\documentclass[12pt,preprint]{aastex}

\newcommand{\ks}{$K_s$\ } 
\newcommand{\kss}{$K_s$} 
\newcommand{\kms}{km s$^{-1}$\ } 
\newcommand{\kslim}{$K_{s,lim}$\ }
\newcommand{\kslimm}{$K_{s,lim}$}
\newcommand{\kslimc}{$K_{s,lim,complete}$\ }
\newcommand{\kslimcc}{$K_{s,lim,complete}$}
\newcommand{\s}{$\sigma$\ }
\newcommand{\stwo}{$\sigma_{200}$\ } 
\newcommand{\stwoo}{$\sigma_{200}$} 
\newcommand{\mtwooo}{$M_{200}$\ } 
\newcommand{\mtwoooo}{$M_{200}$} 
\newcommand{\mtwo}{$M_{vir,200}$\ } 
\newcommand{\mtwoo}{$M_{vir,200}$}
\newcommand{\rtwo}{$R_{200}$\ }
\newcommand{\rtwoo}{$R_{200}$}
\newcommand{\lk}{$L_{K_s}$\ }

\newcommand{\lktwo}{$L_{K_s,200}$\ }
\newcommand{\lktwoo}{$L_{K_s,200}$}
\newcommand{\lfk}{LF$_{K_s}$\ } 
\newcommand{\mlk}{$M_{vir,200}/L_{K_s,200}$\ }
\newcommand{\mlkk}{$M_{vir,200}/L_{K_s,200}$}
\newcommand{\mlkkk}{$M_{200}/L_{K_s,200}$\ }
\newcommand{\jmh}{($J-H$)\ } 
\newcommand{\hmk}{($H-K_s$)\ }
\newcommand{\hmkk}{($H-K_s$)}
\newcommand{\jmk}{($J-K_s$)\ }
 
\shorttitle{$K$--band properties of groups}
\shortauthors{Ramella et al.}
\slugcomment{Revised June 01, 2004}

\begin{document}

\title{$K$--band Properties of Well-Sampled Groups of Galaxies}

\author{Massimo Ramella}
\affil{INAF, Osservatorio Astronomico di Trieste, via G.B. Tiepolo 11, I-34131 Trieste, Italy}
\email{ramella@ts.astro.it}
\author{Walter Boschin}
\affil{Dipartimento di Astronomia, Universit\`a di Trieste, via G.B. Tiepolo 11, I-34131 Trieste, Italy}
\email{boschin@ts.astro.it}
\author{Margaret J. Geller}
\affil{Smithsonian Astrophysical Observatory, 60 Garden St, Cambridge, MA
02138}
\email{mgeller@cfa.harvard.edu}
\author{Andisheh Mahdavi}
\affil{Institute for Astronomy, University of Hawaii, 2680 Woodlawn Drive,
Honolulu, HI 96822}
\email{amahdavi@IfA.Hawaii.Edu}
\and
\author{Kenneth  Rines }
\affil{Yale Center for Astronomy and Astrophysics, Yale University, P.O. Box
208121, New Haven, CT 06520-8121}
\email{krines@astro.yale.edu}

\begin{abstract}
We use a sample of 55 groups and 6 clusters of galaxies ranging in
mass from 7\,$\times$10$^{11} M_{\odot}$ to 1.5\,$\times$10$^{15}
M_{\odot}$ to examine the correlation of the \kss--band luminosity
with mass discovered by Lin et al. (2003).  We use the 2MASS catalog
and published redshifts to construct complete magnitude limited
redshift surveys of the groups. From these surveys we explore the IR
photometric properties of groups members including their IR color
distribution and luminosity function.  Although we find no significant
difference between the group $K_s$ luminosity function and the general
field, there is a difference between the color distribution of
luminous group members and their counterparts (generally background)
in the field. There is a significant population of luminous galaxies
with (H-\kss) $\gtrsim$ 0.35 which are rarely, if ever, members of the
groups in our sample. The most luminous galaxies which populate the
groups have a very narrow range of IR color.  Over the entire mass
range covered by our sample, the \ks luminosity increases with mass as
$L_{K_s} \propto M^{0.64\pm 0.06}$ implying that the mass-to-light
ratio in the \kss--band increases with mass. The agreement between
this result and earlier investigations of essentially non-overlapping
sets of systems shows that this window in galaxy formation and
evolution is insensitive to the selection of the systems and to the
details of the mass and luminosity computations.

\end{abstract}

\section{Introduction}

In the low redshift universe, most galaxies reside in groups (Gott and
Turner 1977; Gregory and Thompson 1978; Faber and Gallagher 1979;
Huchra \& Geller 1982; Ramella et al. 1997; Ramella et
al. 1999). Thus, in spite of the difficulty of determining the
dynamical and photometric properties of these often sparse systems,
they have served as a measure of the universal mass-to-light ratio
(Faber \& Gallagher 1979; Ramella et al. 1997; Tucker et al. 2000;
Bahcall et al. 2000; Carlberg et al. 2001).

Early studies of groups of galaxies are based primarily on surveys
drawn from the Zwicky catalog (Zwicky et al. 1961-1968).  Problems
including the small number of observed members, the membership
assignment itself, and non-uniform photometry led to a large spread in
group mass-to-light ratios even if the median was robust to the myriad
observational problems. Groups thus provided one of the routes to an
estimate of the universal mean cosmological mass density,
$\Omega_m$. Because both the systematic and internal random errors in
mass-to-light ratio determination were large, there was little
consideration of either the presence or the impact of group (cluster)
mass-to-light ratios that vary with mass.

As both photometric and redshift surveys have increased in size and
quality, refined analyses of the data have revealed a potential
dependence of the mass-to-light ratio of systems on the system mass
and/or velocity dispersion.  Girardi et al. (2000) and later Girardi
et al. (2002) used heterogeneous data to demonstrate a dependence of
blue mass-to-light ratio on mass, $M/L_B \propto
M^{0.17-0.23}$. Bahcall \& Comerford (2002) derive an analogous
dependence of $M/L_B$ on X--ray temperature which they attribute to
differences in the ages of the stellar population for galaxies in
groups of different mass. The decrease in the fraction of star-forming
galaxies with the mass or velocity dispersion of groups appeared to
support the argument that the variation in mass-to-light ratio with
mass was a population effect (see e.g. Biviano et al. 1997; Koranyi \&
Geller 2002; Balogh et al. 2004).

Recent analyses by Lin et al. (2003 (L03 hereafter), 2004 (L04
hereafter)) of systems of galaxies based on X--ray data for mass
determination and Two-Micron All-Sky Survey (2MASS, Jarrett et
al. 2000) data for luminosity determination suggest a profoundly
different interpretation of the mass dependence of group mass-to-light
ratios. L03 show that $M/L_{K_s} \propto M^{0.31 \pm 0.09}$, steeper
than, but consistent with, the earlier B-band relations.  Rines et
al. (2004) find a similar dependence of \ks mass-to-light ratio on
system mass and/or velocity dispersion in their study of nine very
well-observed clusters of galaxies. Their mass estimates depend on the
dynamics of the cluster galaxy population.

The variation in infrared color with changes in stellar population is
much smaller than the analogous variation in optical bands. Thus, if
the mass dependence were a population effect, one would expect a
shallower \ks relation. L03 and L04 suggest that the dependence of \ks
mass-to-light ratio on mass provides a new window on the galaxy
formation process. They suggest that the dependence results from lower
efficiency and/or efficient disruption of galaxies in massive systems.

In contrast with L03, L04 and Rines et al. (2004), Kochanek et
al. (2003) use 2MASS data to argue that mass-to-light ratios are
essentially independent of system mass, consistent with the historical
perception that the mass-to-light ratios of groups are roughly
independent of the mass of the system. The explanation of the
difference between the L03, L04 and Kochanek et al. (2003) results is
unclear, but the approaches they take to the the problem are very
different. L03 and L04 analyze sets of systems well-observed in the
X--ray. Kochanek et al. (2001) use N-body simulations to guide their
broad statistical analysis based on a matched filter algorithm. They
use dynamical methods and calibration to X--ray data to estimate
masses.

Here we take an approach in between that of L03, L04, and Kochanek et
al. (2003) to investigate the dependence of \kss--band mass-to-light
ratios on the mass of the system. We compile a set of systems
initially selected from a complete redshift survey with subsequent
deeper spectroscopic surveys (Mahdavi et al. 1999; Mahdavi \& Geller
2004). Most of these systems (but not all) have associated extended
X--ray emission (Mahdavi et al. 2000). We use the complete redshift
surveys as a basis for mass estimation. We supplement our sample with
other optically identified systems to enlarge the sample. The
dependence of \kss--band mass-to-light ratio on mass agrees very well
with the results of L03 and L04.

L03 and L04 use statistical background subtraction rather than
redshift surveys to assess system membership. We examine this
procedure by studying the photometric properties of group members and
non-group galaxies. Although we find a substantial color difference
between the two populations, we show that this difference does not
bias the procedure followed by L03 and L04.

We begin our discussion of the \ks properties of groups with a
discussion of the group catalog and the construction of a complete
magnitude limited redshift list for each group using the 2MASS catalog
(Section 2). Section 3 discusses the infrared photometric properties
of groups members.  Section 3.1 is a discussion of the IR colors of
groups members and non-members (generally background). We discuss the
\kss--band luminosity function (LF) of the groups in our sample in
Section 3.2. In Section 4 we investigate the dependence of \ks light
as a function of the mass of the system as determined from the virial
theorem. We compare the results of Section 4 with L03, L04, and Rines
et al.  (2004) in Section 5 and we conclude in Section 6.  Throughout
this paper we use $H_0=100\,h$ km sec$^{-1}$ Mpc$^{-1}$.

\section{The Group Catalog and Group Membership}

The 2MASS extended source catalog (Jarrett et al. 2000; 2MASS)
provides uniform photometry over the entire sky potentially enabling a
uniform comparison of the photometric and dynamical properties of
systems of galaxies (Kochanek et al.  2003; L03; L04).  To obtain
estimates of system mass and \kss--band luminosity, we compile a set
of poor systems which are well-sampled in redshift space.

We select our group and cluster sample from existing catalogs. We use
galaxy redshifts in 39 well-sampled groups (Mahdavi et al.  1999;
Mahdavi \& Geller 2004).  These systems constitute our "core" sample
because they were selected and observed in a homogeneous way.  Groups
in this sample were identified in an unbiased way from complete,
magnitude limited redshift-surveys (CfA2 and SSRS2). Subsequently
Mahdavi et al. (1999) and Mahdavi \& Geller (2004) measured redshifts
to a deeper magnitude limit within a projected radius $R_{search}$ =
1.5 $h^{-1}$ Mpc.  We supplement this sample with 8 groups from
Zabludoff \& Mulchaey (1998) and 14 AWM/MKW poor clusters from Koranyi
\& Geller (2002). Table 1 lists these 61 systems.

These 61 systems are at low redshift (cz $\lesssim$ 12,000 \kms) and
span a three-order-of-magnitude range in mass.  Most of our systems
have extended X--ray emission, certifying their reliability as
physical systems.  Thirty (77\%) of the Mahdavi et al. (1999, 2004)
groups are associated with extended X--ray emission as are 6 (75\%) of
the Zabludoff \& Mulchaey (1998) and 8 (57\%) of the Koranyi \& Geller
(2002) systems.  The groups not associated with extended X--ray
sources may be below the current detection thresholds.

To obtain an estimate of the group luminosity we use the \kss--band 20
mag arcsec$^{-2}$ isophotal fiducial elliptical aperture magnitudes from
2MASS.  

We use thes magnitudes following Jarrett (2003;
the FAQ sheet for the 
2MASS Extended Source Catalog 
(http://spider.ipac.caltech.edu/staff/jarrett/2mass/XSC/
jarrett\_XSCprimer. html) who emphasizes that 
"the isophotal elliptical magnitudes provide accurate colors for 
galaxies of all sizes" while still "capturing most of the integrated 
flux (~80-90\%)"

For each system from Mahdavi et al. (1999) and Mahdavi \&
Geller (2004), we select all galaxies in the 2MASS extended source
catalog (Jarrett et al. 2000) which lie within 1.5 $h^{-1}$ Mpc of the
center listed in Table 1.  For systems observed by Zabludoff \&
Mulchaey (1999) and Koranyi \& Geller (2002) we search the 2MASS
catalog to the radius listed in Table 1.

We match these 2MASS galaxies with the galaxy redshift list for each
group. For redshifts from Mahdavi et al. (1999, 2004), Zabludoff \&
Mulchaey (1999), and Koranyi \& Geller (2002) we take the membership
assignments given by these authors. We searched NED \footnote {The
NASA/IPAC Extragalactic Database (NED) is operated by the Jet
Propulsion Laboratory, California Institute of Technology, under
contract with the National Aeronautics and Space Administration.} for
additional members of each group.  We include additional galaxies as
members if the redshift is within 3 \s (the velocity dispersion within
the limiting search radius in Table 1) of the group mean
redshift. This procedure yields 90 additional redshifts including 29
additional members.  These last redshifts enable us to extend the
completeness limit of each group redshift survey to a fainter limit.

We rank system members according to their \ks magnitude and identify
the faintest magnitude \kslimc for which the group redshift survey is
complete.  Because the subsamples \ks $\leq$ \kslimc are, in some
cases, rather small, we increase the magnitude limit as much as
possible by requiring that at most one galaxy without a redshift is
included within \kslimm.  The inclusion of a single galaxy without a
measured redshift {\it does} produce a substantial gain in the
sampling of 39 groups. With this procedure, our individual group
surveys reach $\sim$ 0.3 magnitudes fainter and we include a total of
200 (101) additional galaxies (members). In six cases we add more than
ten galaxies to individual groups. Our apparent magnitude limits are
in the range 10.83 $\leq$ \kslimcc $\leq$ 13.45, with a large fraction
close to \ks = 13.0 (median \kslimc = 12.85, with inter-quartile range
i.q.r = .22).  The corresponding absolute magnitude limits peak at
$\sim M_{s,lim,complete}$ = -21.3 with i.q.r. = 0.4.
 
We assign the mean redshift of the group to the single galaxy without
spectroscopy and verify that the inclusion/exclusion of this galaxy
from the member list does not alter any of our results significantly.
In the analysis below we use the samples limited to \kslimm.

The physical quantities we investigate are the mass and total
luminosity in the \kss--band within some fiducial radius.  To obtain a
physically meaningful and stable estimate of the radius, we use
\rtwoo, the radius enclosing an overdensity 200 $\rho_{crit}(z)$
(Carlberg et al. 1997), where $\rho_{crit}(z)$ the critical density
for an Einstein - de Sitter universe at redshift $z$.

Our systems are not rich enough for a reliable fit to a model density
profile (e.g. Navarro et al. 1997; NFW).  We thus assume that the
groups are in virial equilibrium and that their mass increases
linearly with the radius, $r$.  Under these conditions (Carlberg et
al. 1997), $R_{200} = \sqrt3\,\sigma(1 + z)^{-3/2}/(10\,H_o)$ where we
compute \s from all the member galaxies within the limiting search
radius (Table 1) irrespective of \ks magnitude of the galaxies.  This
procedure is similar to the one employed by Carlberg et al. (1997) for
clusters.

The velocity dispersion profiles of groups of galaxies vary. Mahdavi
et al. (1999), Mahdavi \& Geller (2004), and Koranyi \& Geller (2002)
show that the velocity dispersion profiles may be rising, falling or
flat.  As a result of these variations, there is, in general, a
difference between \s and \stwoo, the velocity dispersion within
\rtwoo. However, the median difference between \s and \stwo is
negligible: the median relative difference is 4\% with a narrow 4\%
inter-quartile range.  In the worst case (marked ``a'' in Table 1) the
difference is 30\%, in few other cases the difference is about 20\%,
and in all other cases it is much less.

The median \rtwo is $R_{200,median}$ = 0.7 $h^{-1}$ Mpc with an
interquartile range of 0.18 $h^{-1}$ Mpc.  We compute a virial mass
within \rtwoo: \mtwo = 3 G$^{-1}$ \rtwo \stwoo$^2$.  There are five
systems with fewer than five members brighter than \kslim within
\rtwoo. We exclude these systems (marked ``b'' in Table 1) from
further analysis. We also exclude an additional system where \rtwo is
one third of its search radius (marked ``c'' in Table 1).  We retain
four other groups that have \rtwo slightly larger than their search
radius.  The total sample we analyze then contains 55 systems; 35 of
these systems include a single galaxy without redshift. The final
group sample contains a total of 1192 (955) galaxies (members).

\section {The Infrared Properties of Group Members}

The three IR bands of the 2MASS survey allow investigation of the IR
colors and magnitudes of $\sim 1200$ galaxies within the complete
redshift surveys of our 55 groups. In Section 3.1 we examine the color
distribution of galaxies in groups as a function of absolute magnitude
and compare these distributions with the non-members (generally
background galaxies).

\kss--band spectroscopy of nearby star-forming spiral galaxies reveals
a $\sim$20\% contribution to the \kss--band luminosity from 1000 K
dust (James and Seigar 1999).  We thus explore infrared color-color
diagrams for the group and ``field'' galaxies to assess the importance
of extinction and/or dust emission as a contributor to the \kss--band
light from galaxy groups.

In Section 3.2 we consider the constraints our group redshift surveys
place on the group luminosity and we compare the group luminosity
function (GLF) with the LF for the ``general field'' determined by
Kochanek et al. (2001) and Cole et al. (2001).

\subsection {The infrared Color Distributions and Color-Color Diagrams}

Our sample of 55 groups contains 955 group members and 237 non-member
galaxies with magnitudes measured in all three bands, J, H and \kss,
and with \ks $\leq$ \kslimm.  We compute the absolute magnitude in the
\kss--band, $M_{K_s}$, and derive the quartiles of the distribution of
$M_{K_s}$: $Q_1$ = -23.60, $Q_2$ = -22.76, and $Q_3$ = -22.08.  To
examine the color distributions of members and non-members we separate
galaxies into four classes of absolute magnitude ($M_{K_s} < Q_1$,
$Q_1 \leq M_{K_s} < Q_2$, $Q_2 \leq M_{K_s} < Q_3$, and $M_{K_s} \geq
Q_3$ are intervals I, II, III, and IV respectively).  We show below
that K-corrections have a negligible effect on these distributions.

The four panels of Figure 1 show histograms of the \jmk color of
member galaxies (solid line) and of the non-members (dot-dashed line)
in each absolute magnitude bin (thin line).

The most striking features of the histograms are: a) the very narrow
peak of the color histogram of the (intrinsically) brightest member
galaxies in panel I (i.q.r. = 0.02), and b) the marked difference
between the color distributions of these intrinsically luminous member
and non-member galaxies (panel I). The difference between members and
non-members is still apparent in panel II but disappears for the
intrinsically fainter galaxies in panels III and IV.  Low luminosity
non-members are rare in these magnitude limited samples. The
distribution of colors for the entire sample in each absolute
magnitude range (members and non-members) shifts blue-ward for
intrinsically less luminous galaxies. This effect is the same as the
one observed by Cole et al.  (2001) in their analysis of 2MASS
properties of galaxies in a sample extracted from the 2dF redshift
survey.

Figure 2 shows another view of the narrow peak in the \jmk \ color
distribution for members in quartile I as a function of $M_{K_s}$. The
black dots denote member galaxies; the circles denote the
non-members. The symbol size is proportional to the redshift of the
group. Inspection of the Second STScI Digitized Sky Survey (McLean et
al. 2000; DSS) images shows that most of these luminous members are
early type galaxies. All but one group, SRGb037, contribute members to
this high luminosity bin.  This group has average optical properties
(i.e. \s, redshift, number of members), but its brightest member is a
spiral with ordinary infrared colors. This group has not been detected
as an extended X--ray source.

Figure 3 shows the color-color diagram for the class-I galaxies
(crosses represent field galaxies, black dots are members).  The
median \jmh~ color of the members and non-members are coincident,
\jmh$_{median}$ = 0.72, with very similar first and third quartiles:
(0.70,0.73) and (0.68,0.77) for members and non-members respectively.
In contrast, the \hmk \ color of the non-members is significantly
redder than for members.  Members have a median \hmk$_{mem,median}$ =
0.29 with quartiles (0.27,0.31); non-members have~~
\hmk$_{non-mem,median}$ = 0.40 with quartiles (0.35,0.47).  The first
quartile of the \hmk color distribution of non-members is redder than
the third quartile of the \hmk distribution for members.

The four panels of Figure 4 show the redshift distributions of members
(solid line) and non-members (dot-dashed line) in the four magnitude
bins, from I (most luminous 25\%) to IV (least luminous 25\%).  As
expected, the difference in redshift distribution is impressive for
the most luminous quartile and essentially absent for the least
luminous.  In quartile I, the members have a median redshift of
$z_{median}$ = 0.026 with an inter-quartile range i.q.r. = 0.004; for
the non-members, the median redshift is $z_{median}$ = 0.073 with a
much broader distribution than that for the members, i.q.r. = 0.017.
The difference in median redshift decreases as the intrinsic
luminosity decreases. The additional 360 members and 384 non-members
fainter than \kslim show the same behavior.

We can understand the presence of luminous red galaxies among the
non-members by comparing the color-color diagram of Figure 3 with Figure 1
of Hunt et al. (2002) who examine the effect of hot dust
($\thickapprox$ 600 K -- 1000 K) on near infrared colors. The \hmk
color can be red as a result of dust extinction and/or dust
emission. The arrow in Figure 3 shows the reddening vector.
 
Hunt et al. (2002) use $L$--band photometry to separate the effect of
hot dust emission from extinction.  For equal contributions to the
\kss--band luminosity from the quiescent stellar population and hot
emitting dust, Hunt et al.  (2002) compute that the \hmk color of
galaxies can approach \hmk=1.0.  Extinction affects \jmh more than
\hmkk: galaxies with extinction as large as A$_V$ $\simeq$ 5.0 have
\jmh $\simeq$ 1.3 but ``only'' \hmk $\simeq$ 0.6.

Based on theoretical and observed median colors and the spread of the
stellar populations of normal galaxies, galaxies redder than \hmk=0.35
require a contribution to the \kss--band luminosity from dust
extinction/dust emission (Hunt et al. 2002; Hunt \& Giovanardi 1992;
Giovanardi \& Hunt 1988; Fioc \& Rocca-Volmerange 1999).

We translate the limit \hmk = 0.35 to the median redshift,
$z_{median}$ = 0.08, of non-member galaxies by taking the
evolutionary- and K-correction into account (Poggianti 1997).  We
ignore the details of the color transformations between Hunt et al.
standard colors and 2MASS colors.  The few hundredths of magnitude
resulting from color transformations have no substantive effect on our
comparison with Hunt et al. results (see
http://www.astro.caltech.edu/$\sim$jmc/2mass/v3/ transformations/ for
the 2MASS transformations).

In spite of the differences in redshift distribution, galaxy evolution
and K-correction make a negligible contribution to the color
difference between luminous members and non-members. The brightest
members have infrared colors typical of ellipticals (\jmk
$\thickapprox$ 0.9). In the models of Poggianti (1997) the typical
correction for bandwidth and evolution is $\Delta(J-K_s) \lesssim
0.05$ and the maximum $\Delta(J-K_s) \lesssim 0.1$ for the reddest and
most distant galaxies. These color corrections can not bring the color
distributions into agreement.

A recent spectroscopic survey of 2MASS objects with \jmk $>$ 1.2 and
\ks $<$ 15 shows that 6.3$\pm$0.9\% of the objects are AGNs (Francis
et al. 2004). Most of these AGNs are fainter than \ks = 13 and their
average redshift is 0.23, well in excess of the limits of our
background redshift distribution (Figure 4). We conclude that AGNs do
not make a significant contribution to the \hmk $\gtrsim$ 0.45
population in our sample.

Color gradients within galaxies cannot be responsible for the
luminous red background population. The most luminous background galaxies
are typically at $z \simeq $ 0.1; group members are at $z \simeq$ 0.03. 
The ratio of the $(1+z)^4$ cosmological dimming factors 
between background and member galaxies is $\simeq$ 1.3
corresponding to $\simeq 0.3$ mag arcsec$^{-2}$. Thus 
the colors are not computed within a constant physical aperture. Based on
previous investigations (e.g. Peletier et al. 1990;
Terndrup et al. 1994; Jarrett et al. 2003)
this difference has no practical consequences for our 
color analysis because: a) the difference in physical radius is small
(less than 10\%), b) the color gradients at our physical radii 
are very small \jmk $<$ 0.1 for all galaxy types.

Furthermore there are 16 background galaxies in our sample of bright
galaxies that are redder than \jmk = 1.1 and with redshifts in the range
14000 \kms $<$ cz $<$ 19000 \kms. These 16 galaxies are 20\% of 
the population of bright background galaxies redder than \jmk = 1.1. 
Because  the redshift difference between these galaxies and group members 
is small, the varying physical aperture of the isophotal \kss magnitude
cannot be the source of the observed distance/color effect.

Most of our galaxies with \hmk $\gtrsim$ 0.45 probably owe their color
to hot dust emission rather than extinction because none of these
galaxies have a red enough \jmh $\gtrsim$ 1.0.  The substantial
presence of galaxies with emission from hot dust among the non-member
galaxies is a Malmquist-type selection bias.  At the larger redshifts
typical of the luminous non-members, the reddest galaxies are brighter
than the magnitude limit as a result of the contributions from hot
dust emission.  Without the probable contribution from hot dust, 22\%
of the brightest non-member galaxies would not enter our \ks magnitude
limited sample.

The impact of dust on \hmk colors has received little attention in the
literature to date even though the effect is apparent in 2MASS
redshift surveys (e.g. Figure 5 in
http://www.ipac.caltech.edu/2mass/releases/sampler/sampler.html). The
combined effect of extinction and hot dust emission on the \hmk galaxy
color deserves further investigation. A proper study requires
$L$--band observations to discriminate between extinction and dust
emission. It is interesting that very few galaxies with very red \hmk
colors are members of nearby groups; at the bright end of the GLF
essentially all of the galaxies have standard early-type colors. We
conclude that although emission from hot dust does affect the \hmk
colors of some luminous star-forming galaxies, luminous galaxies with
\hmk$\gtrsim 0.35$ are not typical members of groups in the local
universe.

\subsection {The Group Luminosity Function and Group Luminosities}

The groups in our sample contain 955 members with absolute magnitudes,
mostly brighter than $M_{K_s}$ = -21.50.  We investigate the
constraints that this sample places on the GLF and ask whether the GLF
parameters are consistent with the 2MASS field LF derived by Kochanek
et al. (2001). Exploration of the LF parameters is important because
the values of these parameters influence our estimates of total group
luminosities.

The groups in our sample have different completeness limits in
absolute magnitude, and different richnesses. The richnesses are low
(a median of 15 members per group) and absolute magnitude limits are
bright (usually $M_{K_s,lim} \lesssim -21.50$).  It is thus impossible
to determine individual GLFs. The small number of members does not
even allow for robust normalizations necessary to combine our groups
into a total, or composite, LF.

We can however use the total sample to derive some constraints.  The
total number of galaxies in the groups is $N_{mem}$ = 955.  From all
of these objects, we construct a total ``observed'' histogram,
$H_{o}(\Delta_M)$.  The total number of galaxies is $
\Sigma_{bin}H_{o}(\Delta_M) $ = 955, the total number of groups is
$N_{groups}$ = 55, and the bin size is $\Delta_M = 0.2$ mag.  This
histogram is the dotted histogram in Figure 5.

Next we consider a grid in the Schechter (1976) function parameter
space.  At each node of the grid we compute an ``expected'' histogram
$H_{e}(\Delta_M)$.  The grid consists of 50 $\times$ 50 nodes within
the parameter region defined by $ -24.60 < M_{K_s^*} < -22.00$ and $
-1.40 < \alpha < -0.40$.

For the {\it i}-th group, we sample the Schechter function (with the
parameters of the grid-nodes) within the absolute magnitude range
-26.0 $\leq M_{K_s} \leq M_{K_s,lim,i}$, where $M_{K_s,lim,i}$ is the
completeness limit of group $i$.  The sampling of each Schechter
function is extensive enough (we repeat each sampling 1000 times) to
provide a fair representation of the Schechter function itself.  We
then normalize each sample of the Schechter function to the observed
number of galaxies, $N_{mem,i}$, and build $H_{e}(\Delta_M)$ by
summing the $N_{groups}$ samples.

We compare $H_{e}(\Delta_M)$ to $H_{o}(\Delta_M)$ and judge the
agreement with a $\chi^2$ fit.  Figure 5 shows the histogram (thick
solid line) corresponding to the best fit parameters
($M_{K_s}^*,\alpha)_{bf}$ = (-23.55, -0.84).  The inset in Figure 5
shows the 1- and 2-$\sigma$ confidence level contours around the best
fit ($M_{K_s}^*,\alpha)_{bf}$.  Based on the $\chi^2$ value,
$\chi^2_{\nu = 23}$ = 27 , we do not reject the hypothesis that the
Schechter LF is the parent distribution of the observed
luminosities. The value of $M_{K_s}^*$ is close to the Kochanek et al.
(2001) value of -23.4 (and within the 1-$\sigma$ c.l.  contour).
However, the value $\alpha$ = -0.84 is far from $\alpha = -1.1$
(Kochanek et al. 2001) and outside the 2-$\sigma$ c.l. contour.

The high value of $\alpha$ we find is not surprising.  The sampling of
the systems is too shallow to constrain $\alpha$; much fainter limits
are necessary for a proper constraint. We evaluate the necessary depth
below.

For ($M_{K_s}^*,\alpha)$ = (-23.55, -0.84), the group luminosities
are, on average, (8$\pm$ 6)\% fainter than with (-23.4, -1.1).  Most
of the 8\% difference results from the poorly constrained $\alpha$. If
we chose ($M_{K_s}^*,\alpha)$ = (-23.55, -1.1) we obtain luminosities
differing by only (3 $\pm$ 3)\% from those computed with
($M_{K_s}^*,\alpha)$ = (-23.4, -1.1).  Even the 8\% difference is
small compared with other uncertainties and it does not affect the
slope of the relation between \kss--band luminosity and mass (Section
4).

To better understand the problem of constraining $\alpha$, we use the
simulation to assess the magnitude limit we must reach to obtain a
reliable estimate of the parameter $\alpha$ from the sampling of a
``true'' Schechter LF.  We generate a single simulated system by
sampling a Schechter function with ($M_{K_s}^*,\alpha)$ = (-23.4,
-1.1) within a given absolute magnitude limit, $M_{K_s,lim,sim}$.  We
then apply our fitting procedure to this simulated group.

Figure 6 summarizes this experiment.  The contours show the well-known
correlation betweeen $M_{K_s}^*$ and $\alpha$. Furthermore, the
$1-\sigma$ contour around the best fit value ($M_{K_s}^*,\alpha)$ =
(-23.25, -0.92) is quite wide for $M_{K_s,lim,sim}$ = -22 (panel A).
Panel A is particularly relevant to our observed sample because 46 out
of 56 systems have $M_{K_s,lim,i} \leq -22$.  As we push the sampling
of the simulated groups toward fainter values, the best fit values
move closer to the input values and the confidence level contours
become more restrictive.

For $M_{K_s,lim,sim} = -21$ (panel B), the simulation shows that
$\alpha$ still remains poorly constrained even though the input value
is now within $1-\sigma$ c.l.  contour.  For $M_{K_s,lim,sim} = -19$
(panel C), $\alpha$ is better constrained, but the uncertainty of the
fit ``transfers'' to $M_{K_s}^*$. In fact, the uncertainty in
$M_{K_s}^*$ is larger than $\pm 0.5$ mag. For $M_{K_s,lim,sim} = -17$
(panel D) both $M_{K_s}^*$ and $\alpha$ are finally well determined.
$M_{K_s,lim}= -17$ is six magnitudes fainter than $M_{K_s}^*$ and an
enormous observational challenge.

The correlation between $M_{K_s}^*$ and $\alpha$ together with the
poor constraints on $\alpha$ set by insufficiently faint magnitudes
limits emphasize the need for deep samples for the determination of
GLFs. The correlation between the parameters of the Schechter form of
the LF was noted by Schechter (1976) himself and later confirmed
and/or discussed by many authors, including Colless (1989), Lumsden et
al. (1997), De Propris et al. (2003), Andreon (2004), and Ellis \&
Jones (2004). In fact, Andreon (2004) proposes an alternative
definition of $M^*$ that breaks the correlation with $\alpha$.

Existing cluster LFs based on large and/or deep photometric and
spectroscopic surveys have reached farther below $M^*$ with every
passing year.  Lumsden et al. (1997), Valotto et al. (1997), Rauzy et
al. (1998), Garilli et al. (1999) and Paolillo et al. (2001) use a
variety of surveys to reach 2 to 3 magnitudes below $M^*$.  Goto et
al. (2002) and De Propris et al. (2003) use the Sloan Digital Sky
Survey and the 2DF Survey, respectively to probe the cluster LF to
nearly $M^* + 5$. All of these surveys require a substantial
statistical background correction, but these have also improved in the
most recent studies. Christlein \& Zabludoff (2003) use extensive
spectroscopic surveys of a smaller cluster sample, but in one cluster,
A1060, their LF determination reaches magnitudes $\sim M^* + 7$. The
most recent surveys are deep enough to meet the stringent requirements
of determining the Schechter parameters of the cluster LF.

Spectroscopic studies of less rich and/or poorly sampled systems are
more problematic. Flint et al. (2001) discuss methods for sampling the
very faint end of the galaxy LF and Balogh et al. (2001) discuss the
dependence of the $J$--band LF on environment. In both cases there are
challenges in interpreting the photometric data in the absence of
dense spectroscopic data. Surveys of some small sets of groups are
deep and more complete.  An early redshift survey of MKW4 and AWM4
(Malumuth \& Kriss 1986) reaches about M$^* + 3$, but the faint end
slope, $\alpha$, is essentially unconstrained. Mendes de Oliveira \&
Hickson (1991) reach a similar depth in Hickson Compact groups also
failing to constrain the slope.  Zabludoff \& Mulchaey (2000) reach
M$^* + 4.5$ and obtain parameters of the LF consistent with our
choice.

L04 determine the 2MASS LF to $M_{K_s} = -21$ for a sample of well
known systems of galaxies. They analyze 2 samples of 25 systems each,
one sample including their highest mass systems (out of 93), the other
their lowest mass systems.  L04 find $\alpha \simeq -0.8$ for the
composite LF of each sample, in very close agreement with our
results. Figure 2 of L04 shows two LFs with very different slopes,
$\alpha =$ -1.1 and 0.84. Both LFs provide a satisfactory fit to the
data bright-ward of $M_{K_s} = -21$.

A common practice in the computation of GLFs is elimination of the
brightest galaxy from each group before the fit.  The narrow infrared
color range of the brightest galaxies (Section 3.1) gives some
physical justification for this approach. Eliminating the brightest
galaxies from the fit, we obtain ($M_{K_s}^*,\alpha)_{bf}$ = (-22.95,
-0.54) with a $\chi^2$ small enough to accept the hypothesis that the
Schechter LF accounts for the observed data.  The $2-\sigma$
c.l. contour we obtain without the brightest galaxies does not include
the best fit parameters we obtain from the entire sample of group
members. Nonetheless, the parameters (-22.95, -0.54) used to compute
the total luminosity of groups including the brightest galaxy lead to
an underestimate of the total luminosity of only $\sim$5\%. The impact
of the change in the parameters of the LF on the total luminosity is
small.  However, the brightest galaxy itself typically accounts for
about 40\% of the group luminosity and omitting it from the summed
group luminosity has an obviously large effect.

We conclude that a) a significant variation in $\alpha$ only leads to
a 10\% difference in total luminosities, b) $\alpha$ is poorly
constrained, c) the best fit value $M_{K_s}^* = -23.55$ is within one
bin-width, $\Delta_M = 0.2$ mags,\ of the Kochanek et al. (2001)
value, and d) elimination of the first-ranked groups members does not
change the GLF parameters $M_{K_s}^*$ and $\alpha$. Omission of the
first-ranked galaxy from the observed total group luminosity does have
a substantial effect.  We conclude Kochanek et al. (2001) LF is a
reasonable choice for computation of total group luminosities.

Using the Kochanek et al. (2001) LF parameters ($M_{K_s}^*,\alpha)$ =
(-23.4, -1.1), we integrate the \lfk to $M_{K_s} = -19.5$,
corresponding to the intrinsically least luminous galaxies at the
relevant \kslimm.  We normalize the LF with the observed number of
group members $N_{obs}$ within \rtwo and brighter than $L_{K_s,lim}$,
the luminosity corresponding to \kslim at the mean redshift of the
group:

\begin{equation}
\phi^{*}=\frac{1}{N_{obs}}\int_{L_{K_s,lim}/L_{K_s}^{*}}^{+\infty}\,t^{\,\alpha}\,e^{-t}\,dt
\label{eq1}
\end{equation} 

We then sum the luminosities $L_{K_s,i}$ of the observed members
(including the single galaxy without a redshift) and use the
normalization of equation \ref{eq1} to extrapolate each group
luminosity to the fixed limit $L_{K_s,min}$ corresponding to $M_{K_s}
= -19.5$

\begin{equation} 
L_{{K_s}}=\sum_{i=1}^{N_{obs}} L_{K_s,i} + \phi^{*}L_{K_s}^{*}\int_{L_{K_s,min}/L_{K_s}^{*}}^{L_{K_s,lim}/L_{K_s}^{*}}\,t^{\,\alpha+1}\,e^{-t}\,dt 
\end{equation} 

We note that the integration of the \lfk to 
the common limit  $M_{K_s} = -19.5$ corresponds to an extrapolation
of the observed luminosity of only 10\%-20\% for most of
our systems.

Because not all galaxies without redshifts are real members, we
slightly overestimate the total luminosity for a number of groups. Of
course, a few apparent members {\it with} redshifts may also be mere
superpositions (c.f. Cen 1997). Because these potential non-member
galaxies are not luminous, these effects are small, typically only a
few percent.

Following L03, we finally correct the total luminosities by a factor
1.2 to account for the systematic underestimation of the total light
of galaxies with the 2MASS isophotal magnitudes (Kochanek et
al. 2001).  Table 1 lists masses and corrected luminosities for all of
the systems together with their errors.  For each group, we derive the
error in \mtwo from the distribution obtained with 1000 bootstrap
re-samplings of the redshifts.  For the error in \lk we use the
jackknife re-sampling because, in some cases, repeated samplings of
the brightest galaxy lead to unrealistic luminosities.

\section{The Group K--band Mass-Luminosity Relation}

Although we use the light from galaxies to trace the mass distribution
in the universe, the details of the relationship between mass and
light remain poorly understood from both the theoretical and
observational points of view.  From the observational point of view,
the relation between the mass of an individual galaxy and its
luminosity is affected by current star formation and by the star
formation history. Infrared bands are less affected by current star
formation than optical bands (Gavazzi et al. 1996; Zibetti et al.
2002; Jarrett et al. 2003). Here we 
examine the behavior of \kss--band light as a tracer of the mass 
in systems of galaxies. Emission from the old stellar 
population dominates the \kss--band light in groups of galaxies.

We first examine the relationship between \kss--band light and
mass. L03 and L04 explore the relations $L_{K_s,500}$ vs $M_{500}$
and L$_{K_s,200}$ vs $M_{200}$, respectively, for clusters in the mass
range $2\times 10^{13}\,h^{-1}\,M_{\odot} < M_{200} < 1.2\times
10^{15}\,h^{-1}\,M_{\odot}$.  Our sample extends the mass range to
$10^{12} M_{\odot}$.  In contrast with L03 and L04 who use X--ray
masses and a statistical procedure (not dependent on measurements of
redshifts) to obtain the \kss--band light, we use the virial mass and
a direct measurement of the \kss--band light contributed by the
intrinsically brightest members of the system.

29 (42) of the groups in the ``core''(extended) samples have
associated extended X--ray emission (we mark these groups with an
``X'' in Table 1).  Only a few of them are detected with high enough
signal-to-noise ratio in the X--ray to derive an X--ray mass. To treat
all of the systems homogeneously, we use the virial mass within \rtwo
for all systems.  Our sample is largely independent of those examined
by L03 and L04: one of our ``core'' groups is in the L03 sample,
another one is in the L04 sample, and a further 7 groups in the
extended sample are in the L04 sample.

Figure 7 shows log(\lktwoo) vs log(\mtwoo) for the ``core'' sample of
36 groups from Mahdavi et al. (1999) and Mahdavi \& Geller (2004).  We
use the BCES (Bivariate Correlated Errors and intrinsic Scatter)
estimators for the linear regression analysis (Akritas \& Bershady
1996: http://www.astro.wisc.edu/$\sim$mab/archive/stats/stats.html).
We obtain
\begin{equation}
{\rm log}(L_{K_s,200}) = (0.61 \pm 0.08)\,{\rm log}(M_{vir,200}) + (3.53 \pm 1.0)
\label{LMcore}
\end{equation}
and plot the BCES regression line in Figure 7. From here on, masses
and luminosities are implicitly measured in units of solar values.
The slope is slightly flatter than the 0.72$ \pm 0.04$ obtained by L04
for the relation $L_{K_s,200}$ vs $M_{200}$. The difference between
the L04 relation and ours is insignificant according to the Welch test
(Guest 1961).

Figure 8 shows \mtwo and \lktwo for the total sample of 55 groups.  In
this case the BCES regression analysis leads to
\begin{equation}
{\rm log}(L_{K_s,200}) = (0.56 \pm 0.06)\,{\rm log}(M_{vir,200}) + (4.17 \pm 0.87).
\end{equation}

Figure 8 shows the regression line for the total sample (dotted line)
together with L04 regression line for $L_{K_s,200}$ vs $M_{200}$
(dashed line).  Clearly the behavior of \lktwo vs \mtwo for the
extended sample agrees well with the result obtained for the core
sample (thin solid line), again according to the Welch test. In Figure
8 we shade the area between the two extreme estimates of the
regression line for our core sample (Akritas \& Bershady 1996).  All
the regression lines of our various samples fall within the shaded
area. The L04 best fit relation lies very close to (or within) the
borders of this shaded area.

Our results extend the L04 relation to the low mass range $M_{200} < 2
\times 10^{13}$.  Because there are so few systems within this mass
range, we reconsider the special case of the system NRGb045, dropped
from our virial theorem analysis because it has fewer than 5 members
within \rtwoo.  NRGb045 is the only system in our sample which has an
X--ray temperature unavailable to (and not considered by) L04.  For
this system Mahdavi et al. (2004) use a previously unpublished Chandra
observations to compute the X--ray temperature T$_X$ = 0.61$\pm$0.04
keV.  To obtain a mass, we use the relation of Finoguenov et
al. (2001) between T$_X$ and $M_{500}$ scaled from $M_{500}$ to
$M_{200}$ for a NFW profile with concentration $c = 5$. We obtain
log(\mtwoooo) = 13.05.  Finally we use all available members within
1.5 $h^{-1}$ Mpc down to \kslim = 12.61 to derive a total luminosity
log ($L_{K_s,tot}$) = 11.69.  We mark the position of NRGb045 with the
symbol X in Figure 8. Clearly the low mass system NRGb045 provides
further support for the equation (4).
 
The studies of L03 and L04 indicate that the same relation between
\lktwo vs \mtwooo continues to be valid for masses exceeding those we
sample. The details of their analysis differ from ours.  For
consistency across the entire range of system masses, we analyze
recent cluster data from Rines et al. (2003) and Tustin et al. (2001)
using the same approach we apply to the sample of poorer systems. This
approach avoids systematic offsets which might result from different
approaches to mass and/or luminosity estimation.  Table 2 summarizes
the observations and derived quantities for the 5 clusters surveyed by
Rines et al. (2003) and for the cluster surveyed by Tustin et
al. (2001).  Figure 8 shows the Rines et al. (2003) and Tustin et
al. (2001) clusters as black circles.  Their position in the diagram
agrees with the relation defined by the poorer systems. Including
these clusters in the analysis makes a negligible change in the
regression; the logarithmic slope is now
\begin{equation}
{\rm log}(L_{K_s,200}) = (0.64 \pm 0.06)\,{\rm log}(M_{vir,200}) + (3.19 \pm 0.79).
\end{equation}
We represent this relation with a thick solid line in Figure 8.

Figure 9 shows the mass-to-light ratio, \mlkk, as a function of \mtwo
for the expanded sample in Figure 8 including the Rines et al. (2003)
and Tustin et al. (2001) clusters.  We find
\begin{equation}
{\rm log}(M_{vir,200}/L_{K_s,200}) = (0.56 \pm 0.05)\,{\rm log}(M_{vir,200}) - (5.98 \pm 0.88).
\label{MLM} 
\end{equation}
As discovered by L03, \mlk increases for more massive, higher velocity
dispersion systems.

Mass-to-light ratios of galaxies in the NIR 
vary by no more than a factor of
2 over a large range of star formation histories (e.g. Madau et al 1998;
Bell \& de Jong 2001; Bell 2003).  On the theoretical side, a decrease 
of the differences in mass-to-light ratio toward NIR wavelengths
with variations in stellar population 
is predicted by Bruzual \& Charlot (2003).
The observed/expected range of 
variation in \mlk for individual galaxies is
clearly not enough to produce the observed trend of \mlk vs \mtwo 
for groups.

Uncertainty in the dynamical state of groups, and hence in the validity of
the virial mass estimator, may contribute to the scatter  in
Figure 9. The uncertainty in the ``true'' mass resulting from a
reasonable departure
from the assumed dynamical state of groups is, on average,
$\sim $30\% -- 40\% (e.g. Giuricin et al. 1988;  Diaferio et al. 1999).
This uncertainty is unlikely to alter our results significantly.  A change
in mass by a factor of
1.3 without a corresponding change in luminosity would
move any low mass group
only slightly off the M/L relation. 
It is therefore impossible to explain the
two order of magnitude variation of \mlk we observe over the whole \mtwo range
as a result of evolutionary effects on the mass estimates.

Interlopers, possibly included as group members may also contribute to the
scatter, particularly at the low mass end.  However, the uncertainties in the
luminosity that could be caused by interlopers are much smaller than the
corresponding uncertainties produced in the mass (see for example the error
bars in figure 8).

There are potential systematic variations in galaxy properties with the
velocity dispersion of the system which might contribute to this relation.  We
assume a fixed form for the galaxy LF; it is possible that there are systematic
variations particularly at the faint end. If, contrary to our assumption, the
faint end is steeper for more massive systems, the \mlk would be reduced
relative to less massive systems.  There is some observational evidence for a
larger dwarf-to-giant ratio, or equivalently a steeper faint-end slope in
richer systems (Zabludoff \& Mulchaey 2000). This effect however, cannot be
solely responsible for the variation of the mass-to-light ratio we observe.  We
find an increase by a factor fifty over a mass interval of three orders of
magnitudes.  To explain the \mlk dependence within a mass range of only one
order of magnitude, $\Delta log(M_{vir,200})$ = 1, the LF would have to steepen
up to $\alpha \simeq -2.15$ well outside the observed range.

Variation in the galaxy population as a function of the velocity
dispersion might also contribute to the dependence of \mlk on
\mtwoo. Biviano et al. (1997) and Koranyi \& Geller (2002) show that
the fraction of emission-line galaxies increases as the velocity
dispersion decreases.  The color differences between emission- and
absorption-line galaxies are, however, much smaller at infrared than
at optical wavelengths.  We showed in Section 3.1 that in some
galaxies dust emission makes a significant contribution to the
\kss--band luminosity, however these galaxies are remarkably rare
within groups.  Population effects are thus unlikely to make a
significant contribution to the trend discovered by L03 and L04 and
supported here.

Finally the contributions of the extended halo of the brightest
cluster member and/or intracluster light to the total luminosity are
not included in the 2MASS luminosity.  The presence of intracluster
red giant branch stars (Durrell et al. 2002), planetary nebulae
(Ciardullo et al. 1998; Feldmeier er al. 1998; Durrell et al. 2002;
Feldmeier et al. 2003). globular clusters (West et al.  1995; Jord\'an
et al. 2003), diffuse light (e.g. Zwicky 1952; Melnick et al.  1977;
Uson et al.  1991; Bernstein et al. 1995; Gregg \& West 1998; Gonzalez
et al. 2000), and supernovae (Gal-Yam et al. 2003) not associated with
individual cluster members all suggest that stripped material
contributes to intracluster light (Moore et al. 1999; Gnedin 2003).
In rich clusters like those in the Rines et al. (2003) sample, various
estimates indicate that intracluster light might constitute 5-50\% of
the light in the virial regions.

Two recent studies explore contribution of diffuse optical emission to
the total luminosity of groups of galaxies. White et al. (2003)
examine Hickson compact Group 90 and argue that 38\%-48\% of the total
group light belongs to a diffuse component identified with tidal
debris.  Castro-Rodriguez et al. (2003) carried out a narrow band
survey of the Leo I group to limit the number density of planetary
nebulae in the group. They find none and set a stringent upper limit
of 1.6\% on the contribution of diffuse light to the total luminosity
core of the group. As in rich clusters, the limits on the fractional
contribution of diffuse light to the group luminosity have a similar
and wide range from 1.6\%-48\%.

Recent simulations (Murante et al. 2004) indicate that for systems
with masses exceeding 10$^{14}$M$_\odot$, the fraction of stars in
diffused light increases with cluster mass. They suggest that at least
$\sim$10\% of the stars in a cluster may be contributors to the
intracluster light.

We conclude that the population effects on the relation in Figure 9
are small but that intracluster light could complicate the
interpretation of the relation. Two plausible physical interpretations
of this result are: (1) galaxy formation is less efficient in more
massive systems and/or (2) galaxies are destroyed in collisions and
tidal interactions in the more massive systems. In the second case,
the disrupted material might appear as intracluster light which we do
not detect. There are currently no data available which constrain the
fraction of diffuse light as a function of the mass or velocity
dispersion of the parent system.
 
\section{Comparison with Previous Results}

There are four previous analyses of masses and 2MASS luminosities of
samples of systems of galaxies: Kochanek et al. (2003), L03, Rines et
al.  (2004) and L04.  L03, L04, and Rines et al. (2004) all find a
significant increase of \mlkkk with the mass of the system. Kochanek
et al.  (2003) find no increase and perhaps a small decrease.

In comparing our results with L03 and L04, we focus our discussion on
L04; their large sample of 93 clusters supersedes L03.  Furthermore
L04 and L03 give similar results.  The L04 sample and ours probe
overlapping but not coincident regions in the mass -- luminosity
plane. In particular our sample contains more low mass systems and
fewer high mass systems than L04.  The lowest quartile of the
distribution of masses of L04 systems, $M_{Q25} = 13 \times 10^{13}$,
is larger than our highest quartile, $M_{Q75} = 11\times 10^{13}$. The
lowest masses of our sample (about $10^{12}$) are more than one order
of magnitude below L04 lowest masses ($10^{13}$). We expect this
difference between the mass distributions of the two samples because
L04 systems are X--ray selected.  Even our X--ray emitting systems are
not X--ray selected.

We select our initial ``core'' sample from a redshift survey and
subsequently confirm the physical robustness of each system by X--ray
detection; L04 select a sample of X--ray clusters.  L04 estimate the
total luminosity from de-projected and background corrected counts of
galaxies at the position of X--ray emission peaks of the Abell
clusters in their sample.  They derive masses from X--ray temperatures
whereas we use magnitudes and velocities of individual member galaxies
selected in redshift space to derive a dynamical mass.

L04 fit log(\lktwoo) vs log(\mtwoooo) and obtain $d$ log(\lktwoo)/ $d$
log(\mtwoooo) = (0.72 +/- 0.04). We plot this relation in Figure 8
(dashed line). Clearly the L04 relation is indistinguishable from our
regression lines; a Welch test verifies the visual impression. Given
the completely different methods used to estimate masses and
luminosities, Figure 8 demonstrates the robustness of the different
estimates and of the physical result.

In the intermediate mass range spanned mostly by our ``core'' sample,
there are two systems in common with L04. In the entire sample there
are 9 objects in common (marked ``l'' in Table 1). L04 and our mass
estimates differ for these objects: for 4 out of nine objects we find
a lower mass and for the remaining objects we obtain a higher mass.
The differences are typically a factor of 2 and the median mass ratio
is 1.85.  The system luminosities are in good agreement: the median
ratio between L04 and our luminosities is 1.03 and the fractional
differences never exceed 10\%. In computing this ratio we make a
geometric correction that decreases the luminosities in Table 1 by
20\%: Table 1 lists luminosities projected in cylinders with a radius
\rtwo whereas L04 compute luminosities within a sphere. We also take
the different faint magnitude cut-off of L04 into account (another
10\% decrement of the luminosities in Table 1). The median differences
between the mass and luminosity estimates yield a median mass-to-light
ratio larger than L04 by about 50\% for our overlapping systems.  This
difference corresponds to the median uncertainty on our individual
mass-to-light ratios. The bias in the mass (without a corresponding
bias in the luminosity) roughly preserves the logarithmic relation
between mass-to-light ratio and mass we observe.
 
Because of the small number of overlapping systems, it is difficult to
identify the reason for the differences between the mass estimates.
Interlopers in our systems could artificially increase the velocity
dispersion. Another possibility is that X--ray masses may
systematically underestimate the mass of the system (Finoguenov et al.
2001; Girardi et al. 1998).  Simulations by Rasia et al. (2003) also
suggest the presence of a 30\% - 50\% bias in the direction we find
for masses derived from $\beta$-models.  Taking these potential biases
into account could significantly reduce the observed differences
between our masses and those of L04.

The agreement of our luminosity estimates with those of L04 means that
the presence of a population of dusty objects (see Section 3.1) does
not invalidate the statistical background subtraction of L04.
Statistical background subtraction works here because these
intrinsically luminous objects (with \hmk $\gtrsim$ 0.35), are rarely,
if ever, members of the nearby systems in our sample or in the sample
of L04.

Our masses of the 5 systems in common with Rines et al. (2003) are
larger by a median factor 1.5, systematically exceeding the masses
obtained from caustics. By using the same analysis procedure for these
clusters as for the core sample, we include galaxies in the member
list that lie at high barycentric velocity and at relatively large
radii; these galaxies are outside the caustics.  The number of these
galaxies is small, but their effect is rather large.  We also expect
our masses to be larger than those of Rines et al. (2003) because we
do not correct for the surface term in the virial theorem (Carlberg et
al. 1996; Girardi et al. 1998).  There is no difference between our
luminosity computations and those of Rines et al. (2003).

L04 find that the exclusion of the brightest member galaxy from each
cluster leads to a steeper logarithmic slope of the \lktwo vs \mtwooo
relation.  As noted in Section 3.2, the narrow infrared color range of
the brightest member galaxies of our sample (Section 3.1) gives some
physical justification for this approach.  We exclude the brightest
galaxy from each group before the fit and find the steeper $d$
log(\lktwoo) / $d$ log(\mtwo) = 0.74 +/- 0.06, consistent with the
trend detected by L04.

L04 also plot log($M_{200}/L_{K_s,200}$) vs log(\mtwoooo) and fit it
with a logarithmic slope $\simeq$ 0.3.  Our slope, 0.56$\pm$0.05 is
significantly steeper (figure 9).  It is also steeper than expected on
the basis of our \mlk vs \lktwo relation. One reason for this apparent
inconsistency is the weighting of errors in the particular estimator
of the regression line we use; the fractional errors in the mass are
much larger than the fractional errors in the luminosity biasing the
slope toward steeper values. Furthermore our error bars do not account
for systematic uncertainties and thus the uncertainty in the slope is
thus probably larger than implied by our estimated internal errors.
The shaded area in Figure 8 shows that different estimators (Akritas
\& Bershady, 1996) of the slopes of \mlk vs \lktwo for the expanded
sample have a large spread of (0.54, 0.74). This range of slopes
yields a range of slopes for log($M_{200}/L_{K_s,200}$) vs
log(\mtwoooo) which overlaps the L04 result.

Like L03, L04, and Rines et al. (2004), our results differ from those
of Kochanek et al. (2003).  L03 briefly comment that, in principle,
their sample and the one built by Kochanek et al. (2003) should yield
similar results but that the L04 estimates of the physical properties
of individual systems is more robust that the corresponding estimates
by Kochanek et al.  (2003).  Our selection of systems is more similar
to the procedure followed by L03 and L04 than to the statistical
approach based on structure formation simulations taken by Kochanek et
al.  (2003).  The independent analyses of 55 systems in our sample, 93
systems in the L04 sample, and the 9 CAIRNS clusters (Rines et
al. 2004) show that the increase of the NIR mass-to-light ratio with
mass appears to be a robust property of systems of galaxies with
masses ranging from 7$\times$10$^{11}$ to 1.5$\times$10$^{15}$.

\section {Conclusion}

We use a sample of 55 groups and 6 clusters of galaxies ranging in
mass from 7$\times$10$^{11}$ to 1.5 $\times$10$^{15}$ to examine the
correlation of the $K_s$--band luminosity with mass discovered by L03
and further investigated by L04 and Rines et al. (2004). We use
complete redshift surveys of the 55 groups to explore the IR
photometric properties of groups members including their IR color
distribution and LF.

Although we find no significant difference between the $K_s$--band GLF
and the general field determination by Kochanek et al. (2001), we do
find a difference between the color distribution of luminous group
members and their counterparts (generally background) in the
field. There is a significant population of luminous galaxies with
(\hmk $\gtrsim$ 0.35) which are rarely, if ever, members of the groups
in our sample. The most luminous galaxies which populate the groups
have a very narrow range of IR color.

Although we select and analyze our group sample with approaches
completely different from those taken by L03 and L04, we find nearly
the same dependence of \lktwo on \mtwoooo. The mass-to-light ratio of
groups increases with the mass of the system. Out of the 55 groups
plus 6 clusters we analyze, only 9 systems overlap with the analyses
of L04.

We conclude, as have previous investigators of this issue, that galaxy
formation is suppressed or galaxy disruption is enhanced in more
massive systems. If disruption is the dominant process which accounts
for the dependence of mass-to-light ratio on mass, more massive
systems should harbor relatively more diffuse light.  Recent
simulations give some support to this proposal (Murante et al. 2004).

Neither our analysis nor that of L03 and L04 takes intracluster light
into account. There are no data which set interesting limits on
intra-system light as a function of system mass.  These challenging
observations would be an important contribution to the understanding
of galaxy formation and evolution in galaxy systems.

\acknowledgments

We thank Scott Kenyon and Michael Kurtz for incisive discussions
throughout the course of this work.\\
We thank the anonymous referee for comments that helped us to
improve the paper.\\
This work is partially supported by the Italian Ministry of Education,
University, and Research (MIUR, grant COFIN2001028932 ``Clusters and
groups of galaxies, the interplay of dark and baryonic matter''), by
the Italian Space Agency (ASI), and by INAF (Istituto Nazionale di
Astrofisica) through grant D4/03/IS.\\
The research of MJG and AM was supported in part by Chandra Grant
G02-3179A. AM is a Chandra Fellow.\\
This research makes use of the NASA/IPAC Extragalactic Database
(NED) which is operated by the Jet Propulsion Laboratory, California
Institute of Technology, under contract with the National Aeronautics
and Space Administration.\\
This publication also makes use of data products from the Two Micron All
Sky Survey, which is a joint project of the University of
Massachusetts and the Infrared Processing and Analysis Center, funded
by the National Aeronautics and Space Administration and the National
Science Foundation.\\
This publication makes use of data from the Digitized Sky Survey,
which was produced at the Space Telescope Science Institute under
U.S. Government grant NAG W-2166. The images of these surveys are
based on photographic data obtained using the Oschin Schmidt Telescope
on Palomar Mountain and the UK Schmidt Telescope. The plates were
processed into the present compressed digital form with the permission
of these institutions.

{} 

\newpage
\begin{deluxetable}{l c c c c c c c}
\tablecolumns{10}
\tablewidth{0pc}
\tabletypesize{\scriptsize}
\tablecaption{Basic data for 61 groups in the sample.}
\tablehead{

\colhead{{Group\ ID}} & \colhead{$\alpha$\ {(J2000)}}  &
\colhead{$\delta$\ {(J2000)}}  &  
\colhead{log$_{10}$(M$_{vir,200}$/$h^{-1}$M$_{\odot})$} & 
\colhead{log$_{10}$(L$_{K_s,200}$/$h^{-2}$L$_{K_s,\odot})$}&
\colhead{$r_{\mathrm{search}}$ } &
\colhead{{Comments}} & 
\colhead{{Source}}\\

\colhead{~~} &
\colhead{(h m s)} &
\colhead{($^{\mathrm{o}}\ '\ ''$)}  & \colhead{~~} & \colhead{~~} & 
\colhead{($h^{-1}$ Mpc)} & \colhead{~~}& \colhead{~~} \\

\colhead{(1)} &
\colhead{(2)} &
\colhead{(3)}  & \colhead{(4)} & \colhead{(5)} & 
\colhead{(6)} & \colhead{(7)\tablenotemark{a}}& \colhead{(8)\tablenotemark{b}} 

}

\startdata
{SRGb062      }    &00\ 18\ 22.5 & +30\ 04\ 00 &   13.84$\pm$   0.13 & 12.03$\pm$  0.02& 1.50& {X} &     {M99}  \\  
{SRGb063      }    &00\ 21\ 11.1 & +22\ 18\ 56 &   13.62$\pm$   0.10 & 12.06$\pm$  0.02& 1.50& {X} &     {MG04} \\
{SRGb102      }    &01\ 25\ 55.8 & +01\ 49\ 27 &   13.90$\pm$   0.11 & 11.97$\pm$  0.06& 1.50& {X} &     {MG04} \\
{N664}             &01\ 44\ 02.7 & +04\ 19\ 02 &      -            &      -        & 0.68& {b} &     {ZM88}    \\
{SRGb119      }    &01\ 56\ 13.8 & +05\ 35\ 12 &   13.79$\pm$   0.16 & 11.85$\pm$  0.08& 1.50& {X} &     {M99}  \\  
{SRGb145      }    &02\ 32\ 28.6 & +00\ 56\ 11 &   13.70$\pm$   0.17 & 11.78$\pm$  0.02& 1.50&     -      &     {MG04}  \\
{SRGb149      }    &02\ 38\ 43.8 & +02\ 01\ 11 &   13.90$\pm$   0.12 & 12.09$\pm$  0.04& 1.50&     -      &     {MG04}  \\
{SRGb155      }    &02\ 50\ 19.2 & +00\ 45\ 11 &   14.48$\pm$   0.17 & 11.96$\pm$  0.04& 1.50& {X} &     {MG04}  \\
{AWM7         }    &02\ 54\ 27.5 & +41\ 34\ 44 &   14.71$\pm$   0.06 & 12.40$\pm$  0.01& 1.50& {l, X} &     {KG02} \\     
{SRGb158      }    &02\ 55\ 09.9 & +09\ 16\ 43 &   13.55$\pm$   0.16 & 11.89$\pm$  0.04& 1.50& {X} &     {MG04}  \\
{N2563        }    &08\ 20\ 24.4 & +21\ 05\ 46 &   13.57$\pm$   0.11 & 11.82$\pm$  0.03& 0.62& {l, X} &     {ZM98} \\     
{NRGb004      }    &08\ 38\ 07.3 & +24\ 58\ 02 &   13.40$\pm$   0.17 & 11.96$\pm$  0.02& 1.50& {X} &     {M99}  \\  
{NRGb007      }    &08\ 50\ 29.9 & +36\ 29\ 13 &      -            &      -        & 1.50& {b} &     {M99}    \\
{NRGb025      }    &09\ 13\ 37.3 & +29\ 59\ 58 &   14.05$\pm$   0.16 & 11.83$\pm$  0.05& 1.50& {X} &     {M99}  \\  
{NRGs027      }    &09\ 16\ 20.8 & +17\ 36\ 32 &   13.92$\pm$   0.13 & 12.15$\pm$  0.02& 1.50& {X} &     {MG04}  \\
{AWM1         }    &09\ 16\ 49.9 & +20\ 11\ 54 &   14.31$\pm$   0.10 & 12.20$\pm$  0.02& 1.38&     -      &     {KG02}  \\  
{NRGb032      }    &09\ 19\ 46.9 & +33\ 45\ 00 &   14.03$\pm$   0.12 & 12.12$\pm$  0.03& 1.50& {l, X} &     {M99}  \\  
{MKW1s}            &09\ 20\ 02.1 & +01\ 02\ 18 &      -            &      -        & 0.47& {b} &     {KG02}    \\
{NRGb043      }    &09\ 28\ 16.2 & +29\ 58\ 08 &   13.29$\pm$   0.14 & 11.58$\pm$  0.06& 1.50&     -      &     {M99}  \\  
{NRGB045      }    &09\ 33\ 25.6 & +34\ 02\ 52 &      -            &      -        & 1.50& {b, X} &     {M99}    \\
{NRGb057      }    &09\ 42\ 23.2 & +36\ 06\ 37 &   12.59$\pm$   0.54 & 11.27$\pm$  0.10& 1.50& {X} &     {M99}  \\  
{SS2b144      }    &09\ 49\ 59.9 & --\ 05\ 02\ 48 &   12.87$\pm$   0.24 & 11.50$\pm$  0.07& 1.50&     -      &     {MG04}  \\
{H42          }    &10\ 00\ 13.1 & --\ 19\ 38\ 24 &   13.07$\pm$   0.13 & 11.52$\pm$  0.13& 0.49& {X} &     {ZM98} \\     
{MKW1         }    &10\ 00\ 30.3 & --\ 02\ 58\ 10 &   13.76$\pm$   0.13 & 11.58$\pm$  0.03& 0.94& {X} &     {KG02}  \\  
{NRGs076      }    &10\ 06\ 52.4 & +14\ 27\ 31 &   15.01$\pm$   0.15 & 12.29$\pm$  0.02& 1.50& {X} &     {MG04}  \\
{NRGb078      }    &10\ 14\ 01.8 & +38\ 56\ 09 &   13.77$\pm$   0.11 & 12.00$\pm$  0.04& 1.50&     -      &     {MG04}  \\
{NRGs110      }    &10\ 59\ 09.9 & +10\ 00\ 31 &   14.12$\pm$   0.15 & 12.31$\pm$  0.02& 1.50& {X} &     {MG04}  \\
{NRGs117      }    &11\ 10\ 42.9 & +28\ 41\ 38 &   14.65$\pm$   0.07 & 12.56$\pm$  0.01& 1.50& {l, X} &     {M99}  \\  
{NRGs127      }    &11\ 21\ 34.2 & +34\ 15\ 31 &   13.08$\pm$   0.24 & 12.29$\pm$  0.04& 1.50&     -      &     {M99}  \\  
{SS2b164      }    &11\ 23\ 15.8 & --\ 07\ 51\ 30 &   13.78$\pm$   0.14 & 11.79$\pm$  0.04& 1.50& {X} &     {MG04}  \\
{MKW10        }    &11\ 42\ 23.7 & +10\ 15\ 51 &   12.42$\pm$   0.63 & 11.63$\pm$  0.07& 0.70& {X} &     {KG02}  \\  
{NRGs156      }    &11\ 45\ 33.3 & +33\ 14\ 46 &   13.48$\pm$   0.31 & 11.94$\pm$  0.05& 1.50& {X} &     {MG04}  \\
{MKW4         }    &12\ 04\ 27.2 & +01\ 53\ 43 &   14.24$\pm$   0.11 & 12.16$\pm$  0.03& 1.26& {l, X} &     {KG02}  \\  
{MKW4s        }    &12\ 06\ 38.9 & +28\ 10\ 26 &   14.19$\pm$   0.13 & 12.08$\pm$  0.05& 1.50& {l, X} &     {KG02}  \\  
{NRGb181      }    &12\ 07\ 35.5 & +31\ 26\ 32 &      -            &      -        & 1.50& {b} &     {M99}    \\
{NRGb184      }    &12\ 08\ 55.9 & +25\ 17\ 33 &   13.79$\pm$   0.11 & 11.98$\pm$  0.01& 1.50& {X} &     {MG04}  \\
{AWM2         }    &12\ 15\ 37.6 & +23\ 58\ 55 &   13.60$\pm$   0.14 & 11.69$\pm$  0.04& 0.99&     -      &     {KG02}  \\  
{N4325        }    &12\ 23\ 18.2 & +10\ 37\ 19 &   13.42$\pm$   0.16 & 11.45$\pm$  0.06& 0.95& {X} &     {ZM98} \\     
{H62          }    &12\ 52\ 57.9 & --\ 09\ 09\ 26 &   13.85$\pm$   0.10 & 11.89$\pm$  0.02& 0.56& {l, X} &     {ZM98}  \\  
{NRGs241      }    &13\ 20\ 27.3 & +33\ 12\ 01 &   14.18$\pm$   0.10 & 12.27$\pm$  0.01& 1.50& {X} &     {MG04}  \\
{NRGb244      }    &13\ 23\ 57.9 & +14\ 02\ 37 &   13.43$\pm$   0.14 & 11.85$\pm$  0.06& 1.50& {X} &     {M99}  \\  
{NRGb247      }    &13\ 29\ 25.7 & +11\ 45\ 21 &   13.92$\pm$   0.13 & 12.03$\pm$  0.02& 1.50& {X} &     {M99}  \\  
{NRGb251      }    &13\ 34\ 25.3 & +34\ 41\ 25 &   13.50$\pm$   0.24 & 11.88$\pm$  0.06& 1.50& {a, X} &     {M99}  \\  
{SS2b239      }    &13\ 48\ 51.5 & --\ 07\ 26\ 59 &   13.62$\pm$   0.12 & 11.87$\pm$  0.08& 1.50& {X} &     {MG04}  \\
{MKW5         }    &14\ 00\ 37.4 & --\ 02\ 51\ 29 &   13.28$\pm$   0.15 & 11.71$\pm$  0.05& 0.78&     -      &     {KG02}  \\  
{MKW12        }    &14\ 02\ 48.0 & +09\ 19\ 40 &   13.24$\pm$   0.15 & 11.82$\pm$  0.02& 1.19&     -      &     {KG02}  \\  
{AWM3         }    &14\ 28\ 12.7 & +25\ 50\ 39 &   13.56$\pm$   0.10 & 11.42$\pm$  0.05& 1.34& {X} &     {KG02}  \\  
{NRGb302      }    &14\ 28\ 29.8 & +11\ 29\ 20 &   13.62$\pm$   0.11 & 11.86$\pm$  0.02& 1.50& {X} &     {MG04}  \\
{MKW8         }    &14\ 40\ 42.9 & +03\ 27\ 53 &   13.87$\pm$   0.13 & 12.18$\pm$  0.02& 0.81& {l} &     {KG02}  \\  
{NRGs317      }    &14\ 47\ 05.3 & +13\ 39\ 46 &   13.70$\pm$   0.13 & 11.99$\pm$  0.02& 1.50& {X} &     {M99}  \\  
{N5846        }    &15\ 05\ 47.0 & +01\ 34\ 25 &      -            &      -        & 0.24& {c, X} &     {ZM98}  \\  
{AWM4         }    &16\ 04\ 56.8 & +23\ 55\ 58 &   13.84$\pm$   0.16 & 11.90$\pm$  0.06& 0.56& {l, X} &     {KG02}  \\  
{NRGs385      }    &16\ 17\ 43.9 & +34\ 58\ 00 &   14.39$\pm$   0.08 & 12.28$\pm$  0.01& 1.50& {X} &     {MG04}  \\
{AWM5         }    &16\ 57\ 58.0 & +27\ 51\ 16 &   14.16$\pm$   0.08 & 12.41$\pm$  0.03& 0.73& {X} &     {KG02}  \\  
{H90          }    &22\ 02\ 31.4 & --\ 32\ 04\ 58 &   12.94$\pm$   0.13 & 11.46$\pm$  0.06& 0.33& {X} &     {ZM98}  \\  
{SRGb009      }    &22\ 14\ 46.0 & +13\ 50\ 30 &   13.97$\pm$   0.12 & 11.97$\pm$  0.02& 1.50& {X} &     {M99}  \\  
{SRGb013      }    &22\ 50\ 21.1 & +11\ 34\ 47 &   14.21$\pm$   0.13 & 12.06$\pm$  0.02& 1.50& {X} &     {MG04}  \\
{SRGb016      }    &22\ 58\ 45.9 & +26\ 00\ 05 &   13.71$\pm$   0.10 & 12.13$\pm$  0.04& 1.50& {X} &     {M99}  \\  
{N7582        }    &23\ 18\ 54.5 & --\ 42\ 18\ 28 &   11.83$\pm$   0.49 & 11.30$\pm$  0.07& 0.21&     -      &     {ZM98}  \\  
{SRGb037      }    &23\ 29\ 57.6 & +03\ 40\ 56 &   14.02$\pm$   0.15 & 11.41$\pm$  0.05& 1.50&     -      &     {MG04}  \\
{SS2b312      }    &23\ 47\ 51.6 & --\ 02\ 20\ 16 &   13.36$\pm$   0.22 & 11.70$\pm$  0.04& 1.50& {X} &     {MG04}  \\  
\enddata

\tablecomments{Columns: (1) Name; (2) Right Ascension; (3) Declination; (4) Virial mass within \rtwoo; (5) \kss-band luminosity within \rtwoo; (6) Search radius; (7) Comments; (8) Reference for data source.}

\tablenotetext{a}{Symbols for Column (7): a: $\sigma_{200}=1.3\,\sigma$; b: $<5$ members brighter than \kslim within
\rtwoo; c: $R_{200}\sim 3\,R_{search}$; l: object in common with L04; X: extended X--ray emission.}
\tablenotetext{b}{Symbols for Column (8): M99: Mahdavi et al. 1999;  MG04: Mahdavi \& Geller 2004; ZM98: Zabludoff \& Mulchaey 1998; KG02: Koranyi \& Geller 2002.}


\end{deluxetable}

\begin{deluxetable}{l c c c c c c c}
\tablecolumns{10}
\tablewidth{0pc}
\tabletypesize{\scriptsize}
\tablecaption{Basic data for six Abell clusters.}
\tablehead{

\colhead{{Group\ ID}} & \colhead{$\alpha$\ {(J2000)}}  &
\colhead{$\delta$\ {(J2000)}}  &  
\colhead{log$_{10}$(M$_{vir,200}$/$h^{-1}$M$_{\odot})$} & 
\colhead{log$_{10}$(L$_{K_s,200}$/$h^{-2}$L$_{K_s,\odot})$}&
\colhead{$r_{\mathrm{search}}$ } &
\colhead{{Source}}\\

\colhead{~~} &
\colhead{(h m s)} &
\colhead{($^{\mathrm{o}}\ '\ ''$)}  & \colhead{~~} & \colhead{~~} & 
\colhead{($h^{-1}$ Mpc)} & \colhead{~~}\\ 

\colhead{(1)} &
\colhead{(2)} &
\colhead{(3)}  & \colhead{(4)} & \colhead{(5)} & 
\colhead{(6)} & \colhead{(7)\tablenotemark{a}} 

}

\startdata
A496              &04\ 33\ 35.2 & --\ 13\ 14\ 45 &   14.61$\pm$   0.06    &  12.66$\pm$   0.01 & 1.50&     R03  \\  
A539              &05\ 16\ 32.1 &    +06\ 26\ 31 &   14.67$\pm$   0.06    &  12.52$\pm$   0.01 & 1.50&     R03  \\
A1367             &11\ 44\ 36.2 &    +19\ 46\ 19 &   14.70$\pm$   0.06    &  12.65$\pm$   0.01 & 1.50&     R03  \\
A1644             &12\ 57\ 11.6 & --\ 17\ 24\ 34 &   15.19$\pm$   0.06    &  13.03$\pm$   0.04 & 1.50&     T01   \\
A1656\ (Coma)     &12\ 59\ 31.9 &    +27\ 54\ 10 &   15.02$\pm$   0.03    &  13.01$\pm$   0.01 & 1.50&     R03  \\  
A2199             &16\ 28\ 39.5 &    +39\ 33\ 00 &   14.68$\pm$   0.06    &  12.71$\pm$   0.01 & 1.50&     R03  \\
\enddata

\tablecomments{Columns: (1) Name; (2) Right Ascension; (3) Declination; (4) Virial mass within \rtwoo; (5) \kss-band luminosity within \rtwoo; (6) Search radius; (7) Reference for data source.}

\tablenotetext{a}{Symbols for Column (7): R03: Rines et al. 2003; T01: Tustin et al. 2001.}


\end{deluxetable}

\clearpage

\begin{figure}
\plotone{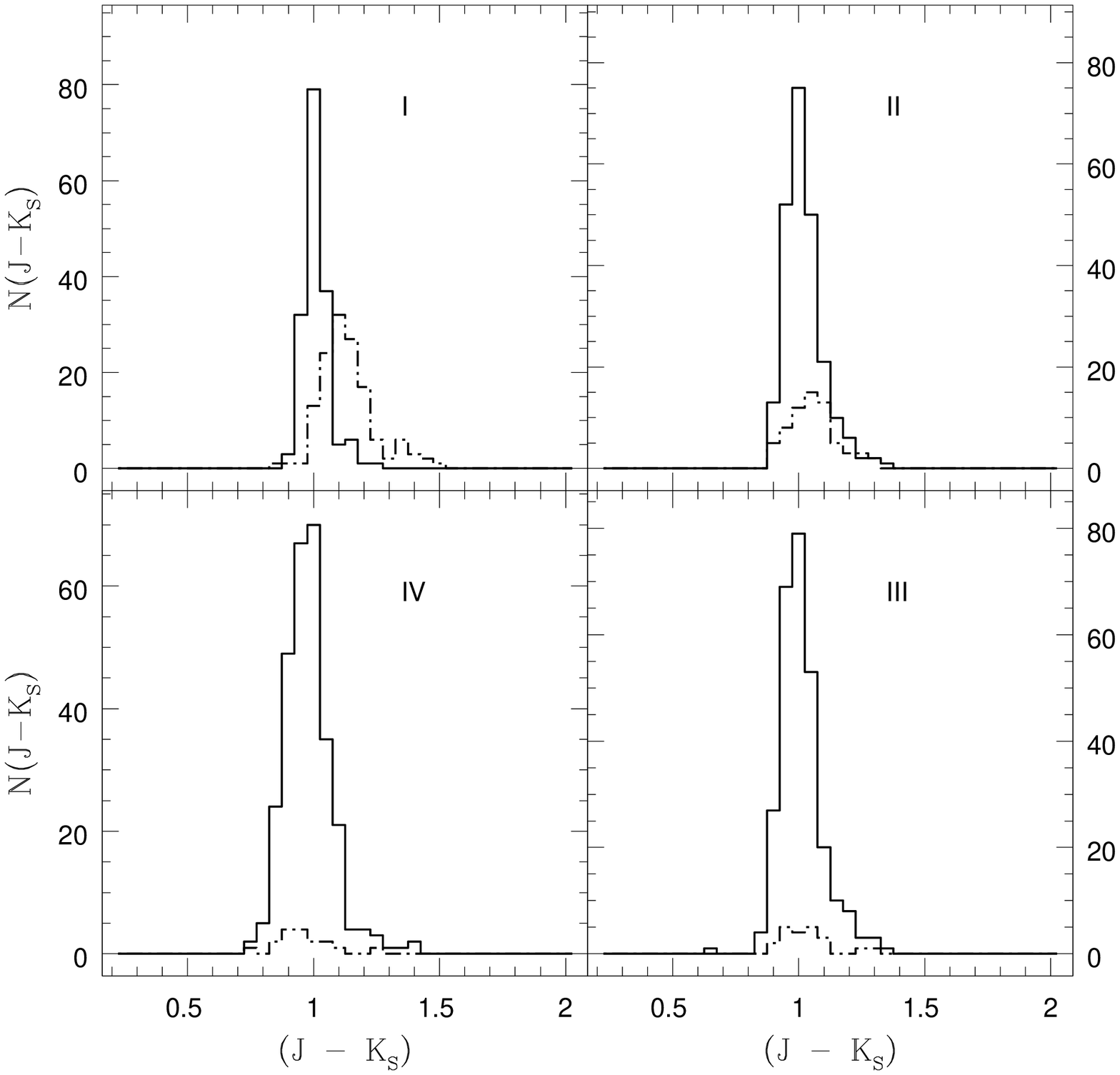}
\caption{\jmk color distribution of members (solid line) and non-members (dot-
dashed line). The four panels (I to IV) are for galaxies of decreasing
luminosity: from galaxies brighter than the first quartile of the 
absolute magnitude  distribution (class I) to galaxies fainter than the third 
quartile (class IV).} 
\label{fig1}
\end{figure}

\begin{figure}
\plotone{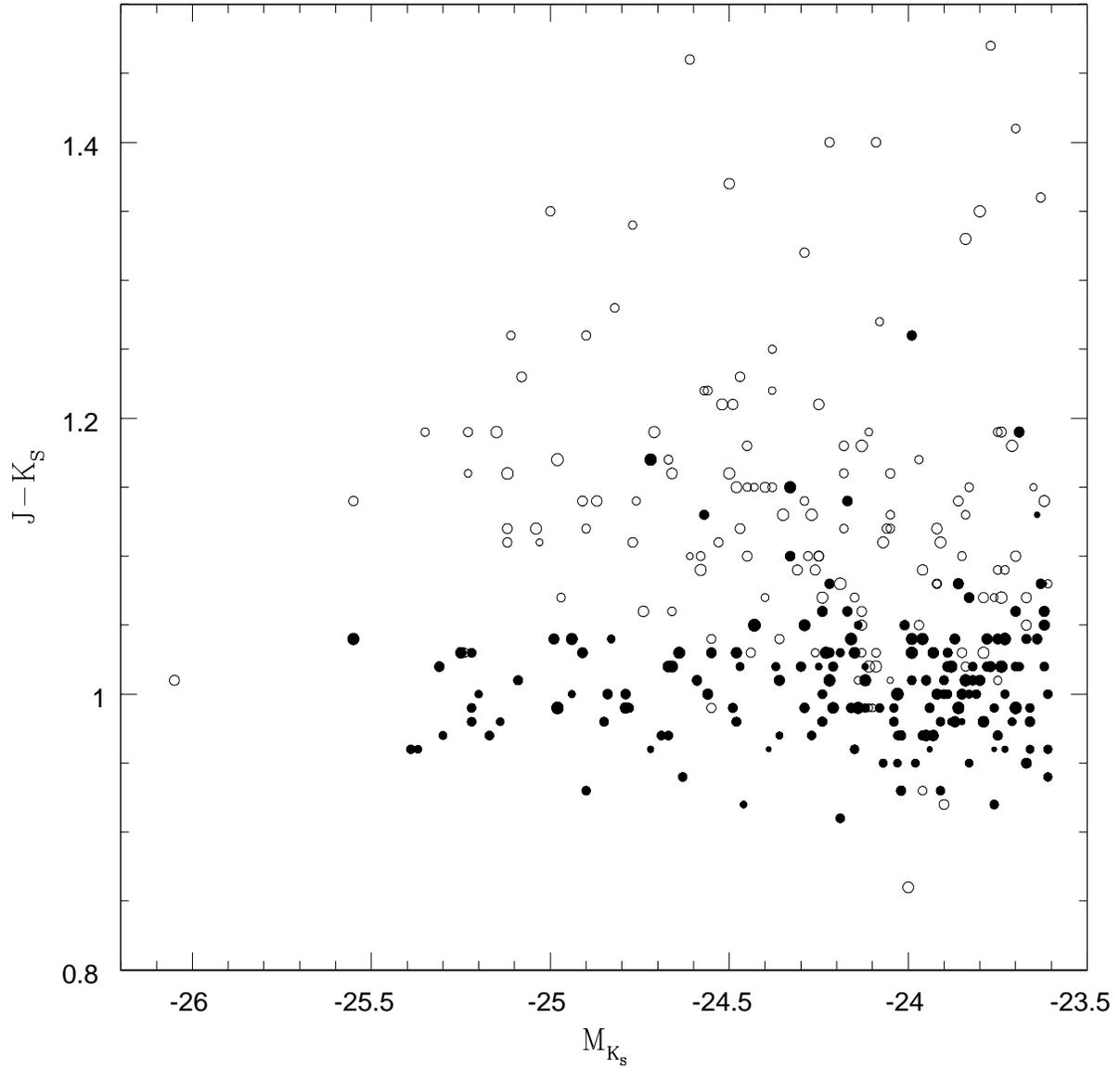}
\caption{\jmk color vs absolute magnitude of members (filled circles) 
and non-members (empty circles). Galaxies are those brighter than 
the first quartile of the  absolute magnitude  distribution (class I).
The sizes of the circles are proportional to the redshifts (larger circles
represent more distant objects).} 
\label{fig2}
\end{figure}

\begin{figure}
\plotone{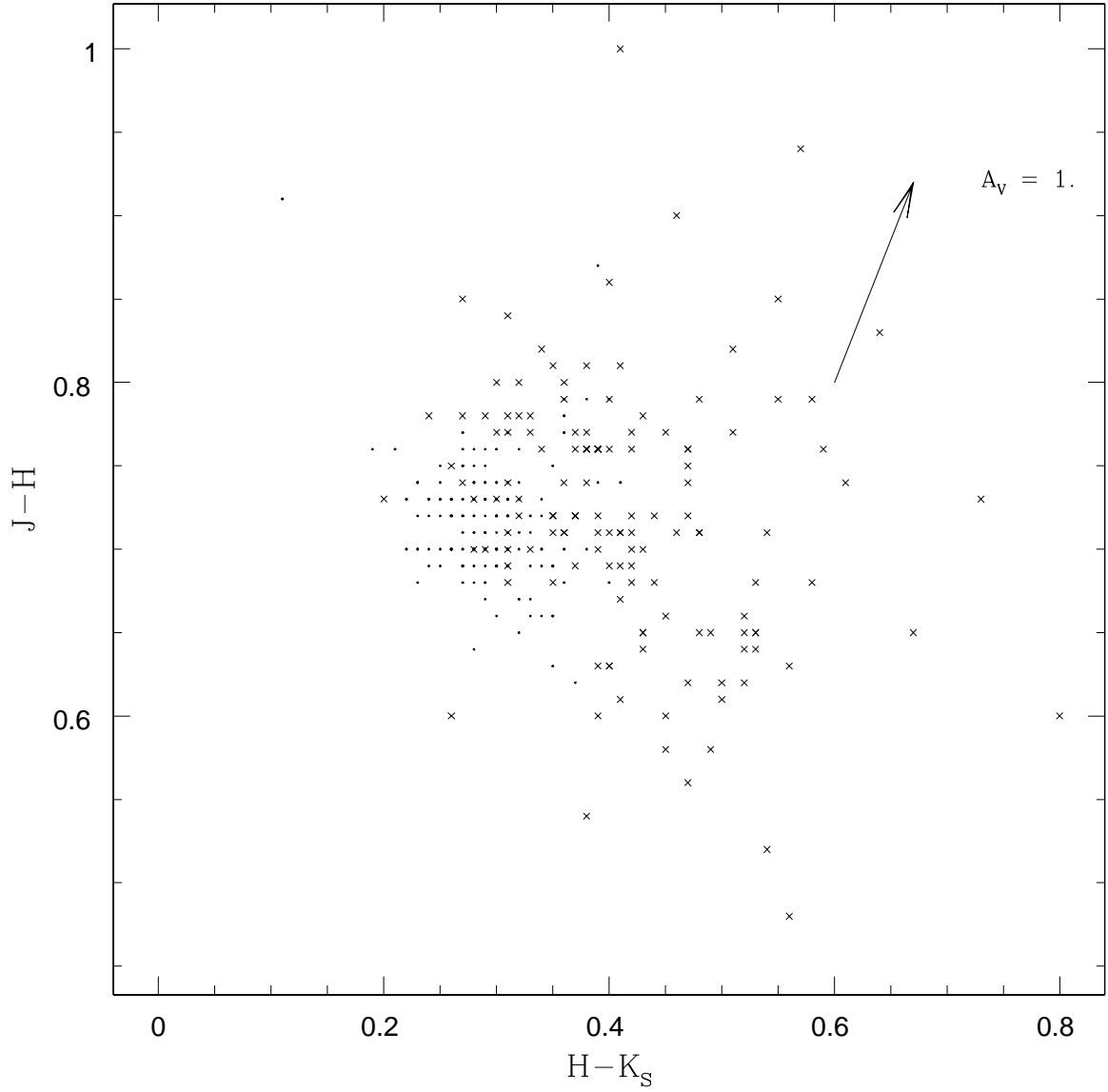}
\caption{Color - color diagram for the brightest galaxies (class I). Dots
represent members, crosses are non-members. The arrow represents the
reddening vector.} 
\label{fig3}
\end{figure}

\begin{figure}
\plotone{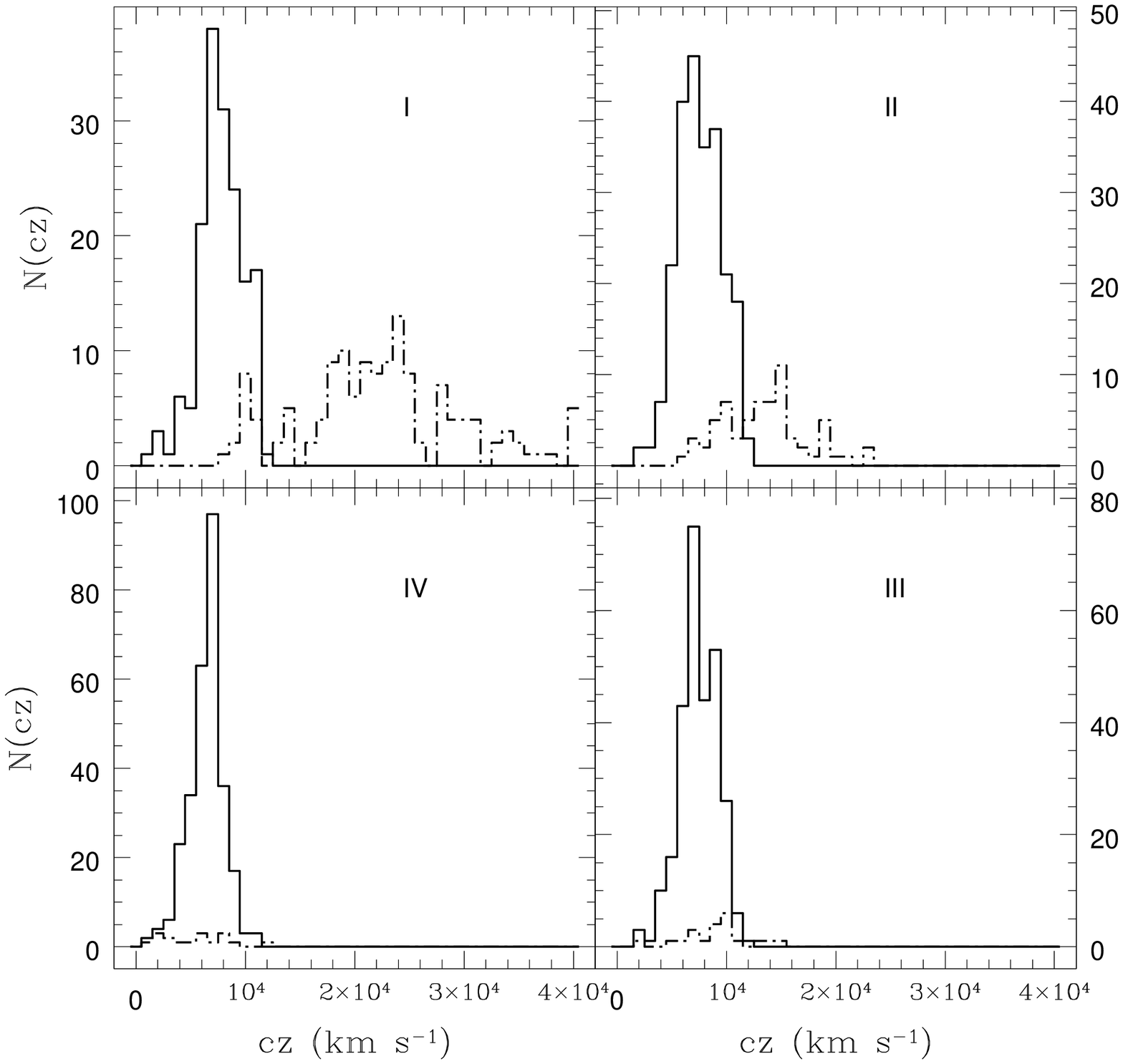}
\caption{Redshift distribution of members (solid line) and non-members (dot-
dashed line). The four panels (I to IV) are for galaxies of decreasing
luminosity: from galaxies brighter than the first quartile of the 
absolute magnitude  distribution (class I) to galaxies fainter than the third 
quartile (class IV).}
\label{fig4}
\end{figure}

\begin{figure}
\plotone{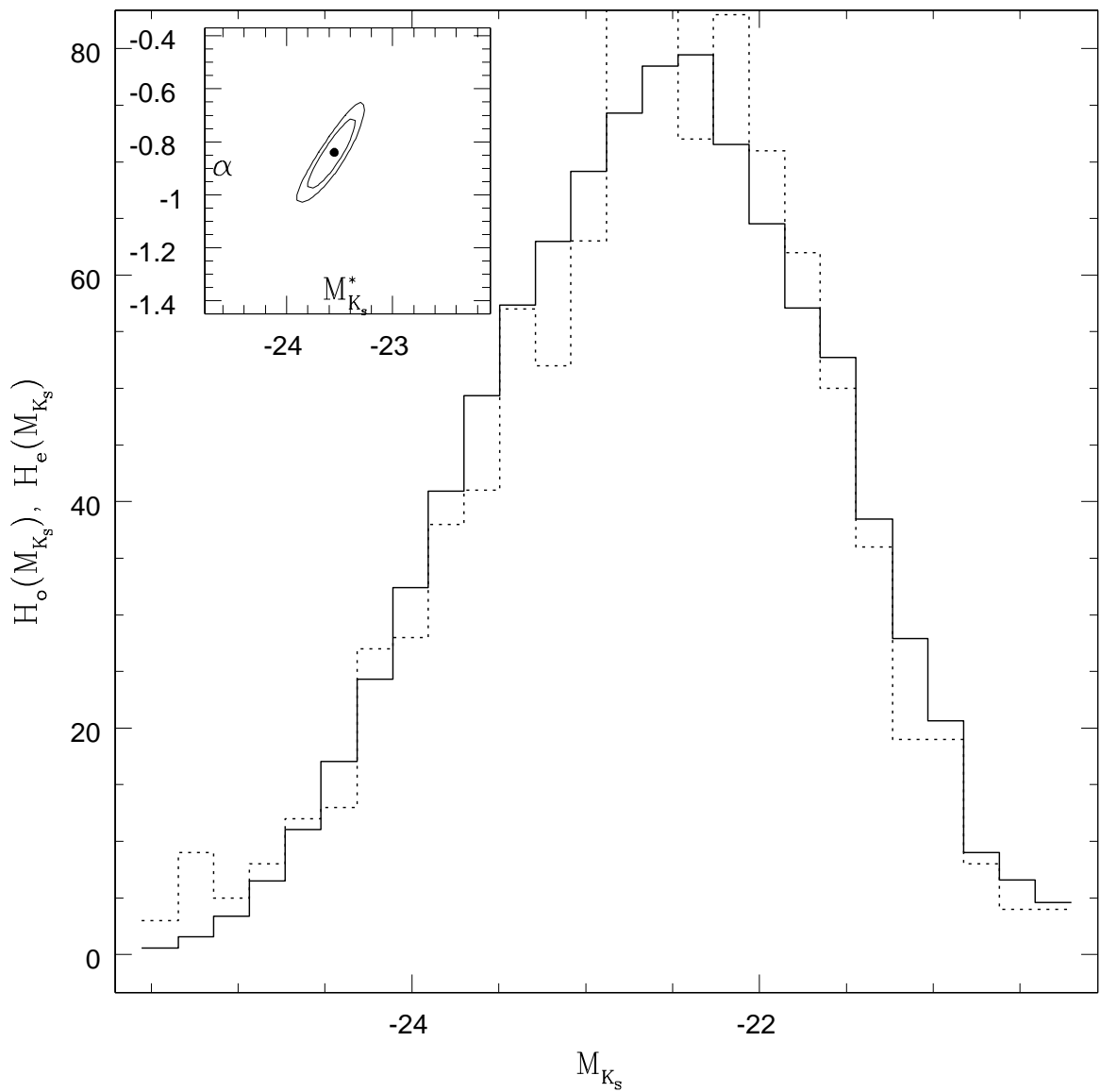}
\caption{Total ``observed'' histogram (dotted line) of absolute
magnitudes of members, $H_{o}(\Delta_M)$, and total ``expected''
histogram $H_{e}(\Delta_M)$ (solid line) computed for a Schecter LF
with the best fit parameters ($M_{K_s}^*,\alpha)_{bf}$ = (-23.55,
-0.84). These values are marked with a dot in the inset. The inset
also shows 1-$\sigma$ and 2-$\sigma$ c.l. contours.  }
\label{fig5}
\end{figure}

\begin{figure}
\plotone{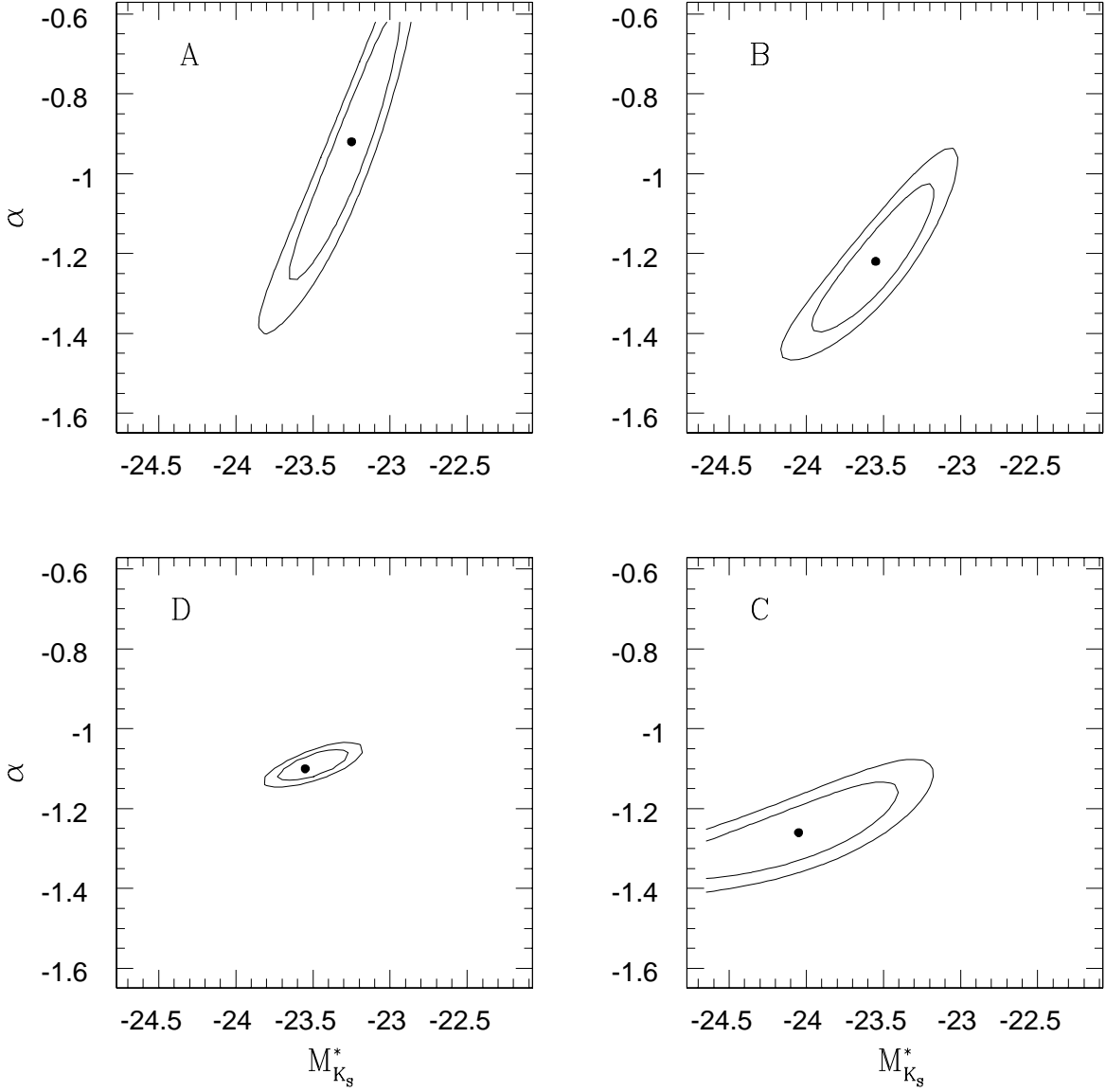}
\caption{ Best fit parameters (dots) and 1-$\sigma$, 2-$\sigma$
c.l. contours obtained for a single simulated system with
($M_{K_s}^*,\alpha)$ = (-23.4, -1.1) within a given absolute magnitude
limit, $M_{K_s,lim,sim}$.  Panels A to D are for $M_{K_s,lim,sim}$ =
-22.0, -21, -19., and -17.0 respectively.  }
\label{fig6}
\end{figure}
\clearpage

\begin{figure}
\plotone{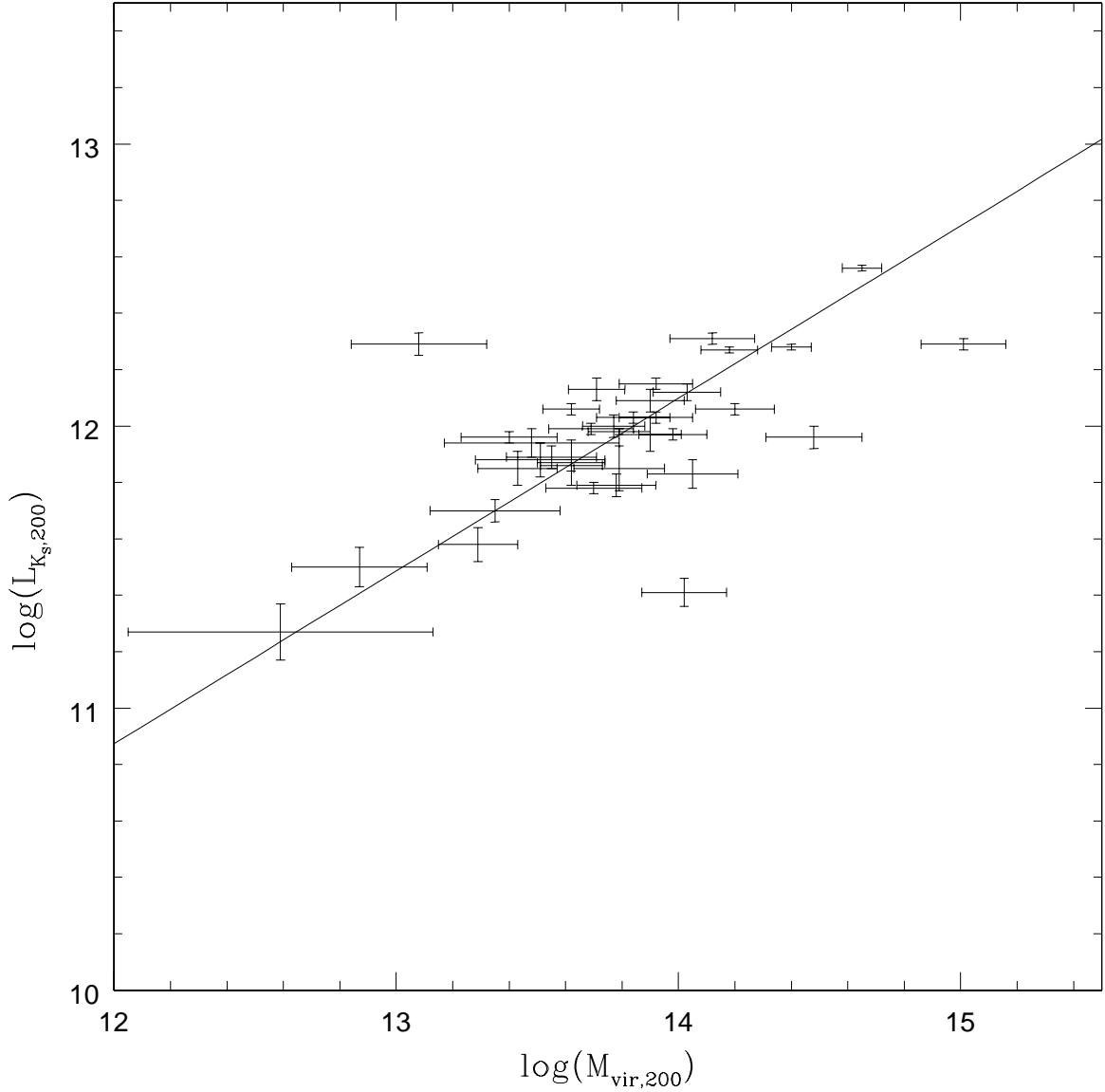}
\caption{log(\lktwoo) vs log(\mtwoo) for the ``core'' sample of 36
groups from Mahdavi et al. (1999) and Mahdavi \& Geller (2004). The
line represents the relation of equation \ref{LMcore}.  }
\label{fig7}
\end{figure}

\begin{figure}
\plotone{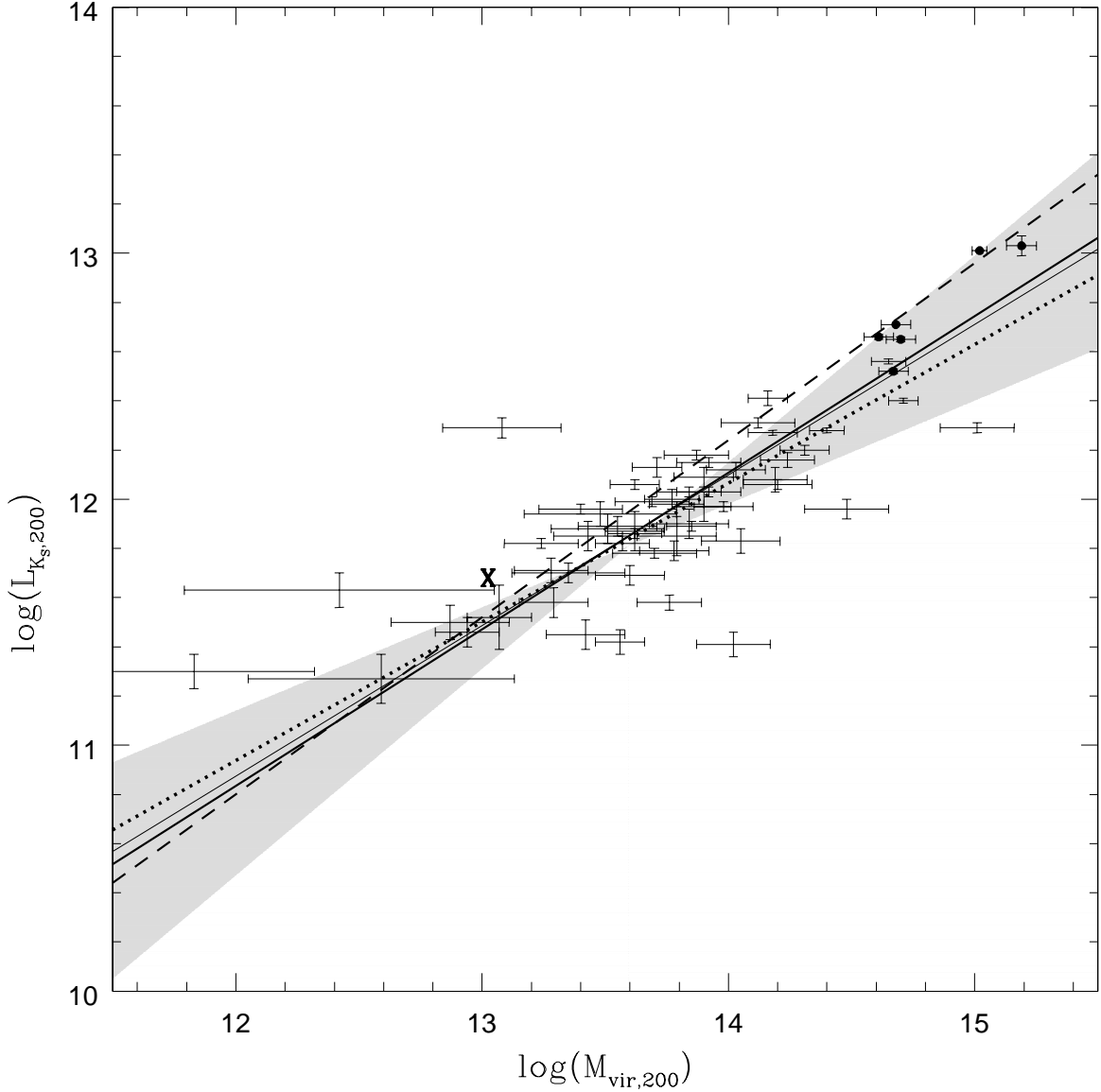}
\caption{log(\lktwoo) vs log(\mtwoo) for the ``expanded'' sample of 55
groups and the 6 clusters of Rines et al. (2003) and Tustin et
al. (2001) (black dots). The lines represent the relations for the
``core'' sample (dotted line), for the ``expanded'' sample (solid
line), and for the sample including both the ``expanded'' sample and
the 6 clusters of Rines et al. (2003) and Tustin et al. (2001) (thick
solid line). The dashed line is L04 relation.  The shaded area
indicates the region between the two extreme estimators of the
relation for the ``core'' sample. The letter ``X'' marks the luminosity and
X--ray mass of NRGb045.}
\label{fig8}
\end{figure} 

\begin{figure}
\plotone{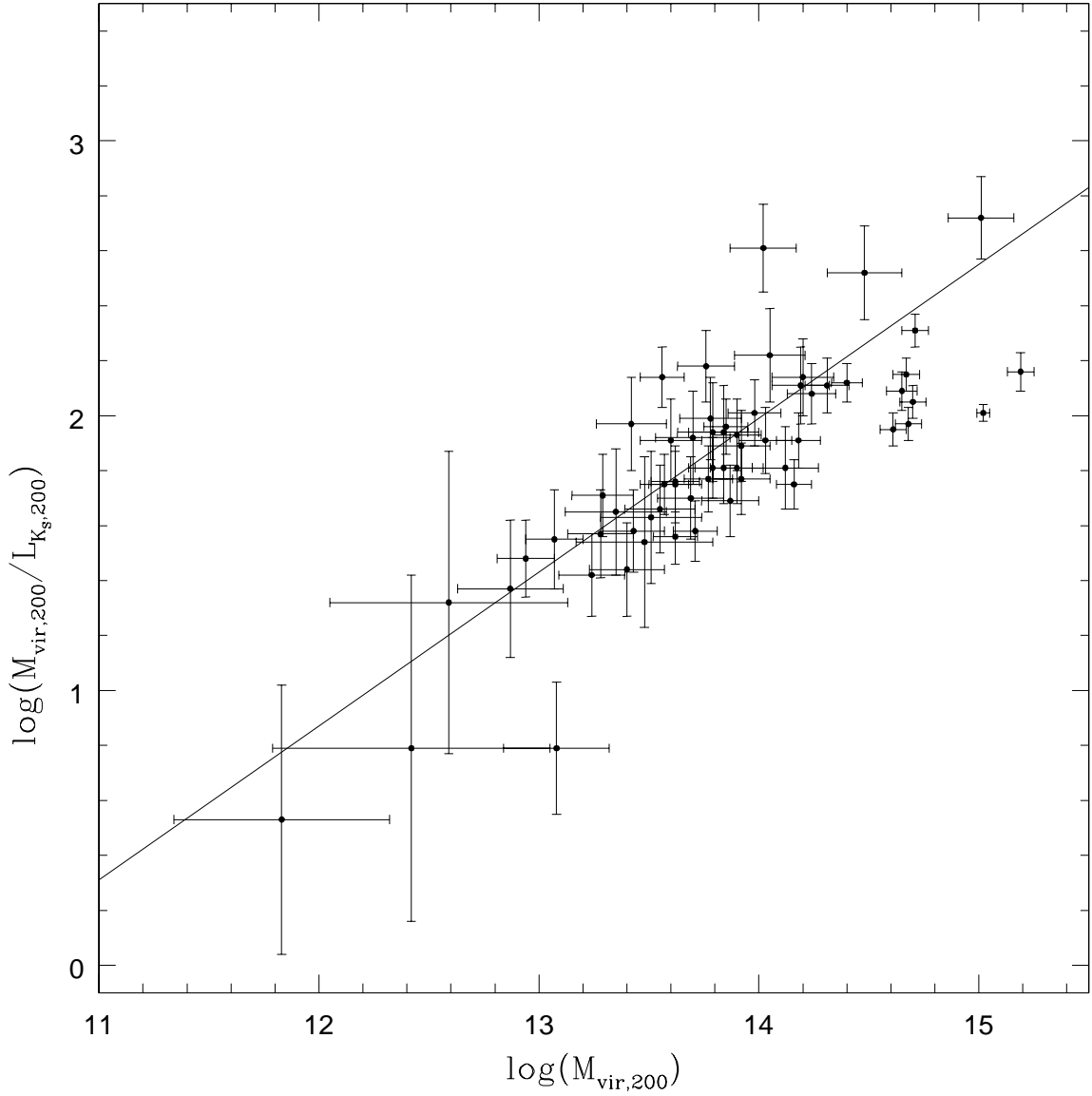}
\caption{log(\mlkk) vs log(\mtwoo) for the ``expanded'' sample of 55
groups and the 6 clusters of Rines et al. (2003) and Tustin et
al. (2001) (black dots). The line represents the relation of equation
\ref{MLM}.}
\label{fig9}
\end{figure}
\end{document}